\documentclass[sigconf]{acmart}

\AtBeginDocument{%
  \providecommand\BibTeX{{%
    \normalfont B\kern-0.5em{\scshape i\kern-0.25em b}\kern-0.8em\TeX}}}

\copyrightyear{2026}
\acmYear{2026}
\setcopyright{cc}
\setcctype{by}
\acmConference[CHI '26]{Proceedings of the 2026 CHI Conference on Human Factors in Computing Systems}{April 13--17, 2026}{Barcelona, Spain}
\acmBooktitle{Proceedings of the 2026 CHI Conference on Human Factors in Computing Systems (CHI '26), April 13--17, 2026, Barcelona, Spain}
\acmPrice{}
\acmDOI{10.1145/3772318.3790536}
\acmISBN{979-8-4007-2278-3/2026/04}

\usepackage{colortbl}
\usepackage{enumitem}
\usepackage{multirow}
\usepackage{csquotes}
\usepackage{verbatim}
\usepackage{subcaption}
\usepackage{graphicx}
\usepackage{soul}

\def\markup{0}
\if\markup1
\newcommand{\rv}[1]{{\leavevmode\color{blue}#1}}
\else
\newcommand{\rv}[1]{#1}
\fi

\def\minor{0}
\if\minor1

\else

\fi

\begin{document}

\title[Multi-Session Study of UX Evaluators Using Conversational AI Agents]{\textit{``It Became My Buddy, But I'm Not Afraid to Disagree''}: A Multi-Session Study of UX Evaluators Collaborating with Conversational AI Assistants}





\author{Emily Kuang}
\affiliation{%
    \institution{York University}
    \city{Toronto}
    \country{Canada}
}
\email{ekuang@yorku.ca}

\author{Ehsan Jahangirzadeh Soure}
\affiliation{%
  \institution{University of Waterloo}
  \city{Waterloo}
  \state{Ontario}
  \country{Canada}
}
\affiliation{
  \institution{Snowflake}
  \city{Toronto}
  \state{Ontario}
  \country{Canada}
}
\email{ehsan.soure@snowflake.com}

\author{Luyao Shen}
\affiliation{%
    \institution{The Hong Kong University of Science and Technology (Guangzhou)}
    \city{Guangzhou}
    \country{China}
}
\email{lshen595@connect.hkust-gz.edu.cn}

\author{Nitesh Goyal}
\affiliation{%
  \institution{Google Deepmind}
  \city{New York City}
  \state{New York}
  \country{USA}
}
\email{teshgoyal@acm.org}

\author{Mingming Fan}
\affiliation{%
    \institution{The Hong Kong University of Science and Technology (Guangzhou)}
    \city{Guangzhou}
    \country{China}
}
\affiliation{
  \city{Hong Kong SAR}
  \country{China}
}
\email{mingmingfan@ust.hk}

\author{Kristen Shinohara}
\affiliation{%
  \institution{Rochester Institute of Technology}
  \city{Rochester}
  \state{New York}
  \country{USA}
}
\email{kristen.shinohara@rit.edu}

\renewcommand{\shortauthors}{Kuang et al.}

\renewcommand{\shortauthors}{}


\begin{abstract} 
AI-assisted usability analysis can potentially reduce the time and effort of finding usability problems, yet little is known about how AI's perceived expertise influences evaluators' analytic strategies and perceptions over time. We ran a within-subjects, five-session study (six hours per participant) with 12 professional UX evaluators who worked with two conversational assistants designed to appear novice- or expert-like (differing in suggestion quantity and response accuracy). We logged behavioral measures (number of passes, suggestion acceptance rate), collected subjective ratings (trust, perceived efficiency), and conducted semi-structured interviews. Participants experienced an initial novelty effect and a subsequent dip in trust that recovered over time. Their efficiency improved as they shifted from a two-pass to a one-pass video inspection approach. Evaluators ultimately rated the experienced CA as significantly more efficient, trustworthy, and comprehensive, despite not perceiving expertise differences early on. We conclude with design implications for adapting AI expertise to enable calibrated human-AI collaboration.

\end{abstract}

\ccsdesc[500]{Human-centered computing~Empirical studies in HCI}
\ccsdesc[500]{Human-centered computing~Natural language interfaces}
\ccsdesc[500]{Human-centered computing~Usability testing}

\keywords{User experience; Usability analysis; Human-AI collaboration; Proactive conversational assistants; Longitudinal study; Novelty effect}

\begin{teaserfigure}
    \includegraphics[width=\textwidth]{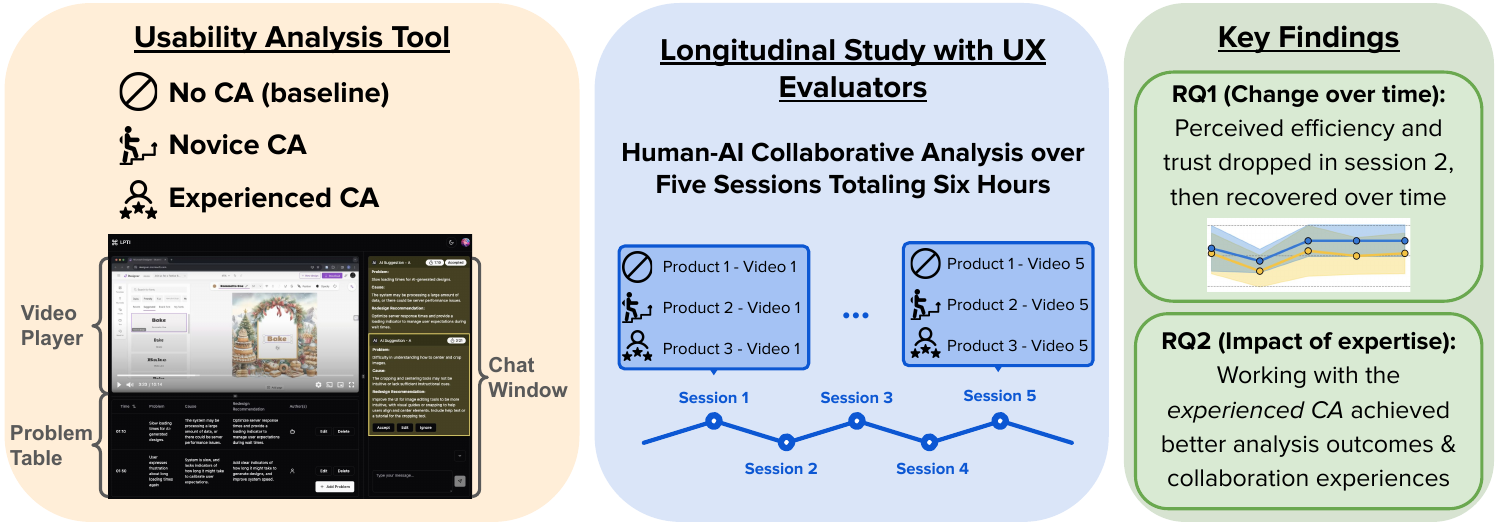}
    \caption{Visual abstract summarizing the study. We developed a Usability Analysis Tool with three conditions: (1) No Conversational Assistant (CA), (2) a Novice CA, and (3) an Experienced CA. In a within-subjects study spanning five sessions, twelve UX evaluators analyzed 15 usability videos across all three conditions. We investigated how their analysis behaviors and perceptions of the CAs changed over time (RQ1), and how the perceived level of CA expertise influenced these outcomes (RQ2).}
    \label{fig:teaser}
    \Description{Visual abstract summarizing the study: The first part shows a Usability Analysis Tool with three conditions: No CA, Novice CA, and Experienced CA. The second part shows a longitudinal study with twelve UX evaluators who analyzed 15 videos under all three conditions. The third part shows the key findings: "RQ1 (Change over time): Novelty effect caused lower perceived efficiency and trust in session 2, which recovered over time; RQ2 (Impact of expertise): Working with the experienced CA achieved better analysis outcomes \& collaboration experiences."}
\end{teaserfigure}
\maketitle

\section{Introduction}

\rv{
Technology is now embedded in nearly every aspect of daily life. As digital interfaces grow more complex, delivering a meaningful and seamless user experience (UX) has become increasingly challenging \cite{norman_definition_1998}. Without good usability, even powerful tools risk going unused. Usability reflects how easily people can use a system, assessed across learnability, efficiency, memorability, error rates, and satisfaction \cite{nielsen_usability_2012}.
Usability testing is central to identifying obstacles to users' goals and generating redesign recommendations. 
It typically involves observing users performing tasks, with sessions often video-recorded for later review \cite{vredenburg_survey_2002, fan_practices_2020, nielsen_usability_1994}. 
These videos help evaluators identify problems by examining users' actions and context, yet reviewing them remains demanding and time-consuming \cite{norgaard_what_2006, chilana_understanding_2010}.
To improve the completeness and reliability of usability evaluations, researchers and practitioners have applied artificial intelligence (AI) and machine learning (ML) techniques to analyze usability recordings \cite{fan_automatic_2020, paterno2017customizable, harms2019automated, jeong2020detecting, yang_re-examining_2020, zimmerman_ux_2020, oztekin_machine_2013, lu_ai_2024}. 
Commercial platforms now offer AI-powered features such as emotion detection and sentiment analysis \cite{uxtesting_uxtesting_2022, usertesting_usertesting_2021}, since frustration and negative sentiment often indicate usability problems \cite{fan_automatic_2020}. 
Academic work similarly analyzes anomalies in interaction logs (e.g., clicks, navigation paths, time on page) to detect issues \cite{paterno2017customizable, harms2019automated, jeong2020detecting, yang_re-examining_2020, zimmerman_ux_2020, oztekin_machine_2013}.
Despite these advances, fully automated approaches remain incomplete, often missing complex or subtle usability problems that require contextual understanding of user expectations and emotional responses \cite{fan_automatic_2020, grigera2017automatic, liu_chatting_2023}. 
This underscores the continued need for human expertise to interpret user behaviors and feedback.

The limitations of fully automated approaches for detecting usability problems have led to increasing interest in \textit{human-AI collaborative analysis}, where AI augments rather than replaces human judgment \cite{lai_towards_2021}. 
Usability pioneer Jakob Nielsen describes AI as a ``free colleague,'' suggesting that the combination of human expertise and AI's scalable processing can exceed the capabilities of either alone \cite{nielsen_what_2024}. 
Prior work has begun to map the landscape of human-AI collaborative usability analysis \cite{fan_vista_2020, fan_human-ai_2022, batch_uxsense_2023, kuang_collaboration_2023, kuang_enhancing_2024}, examining dimensions such as (1) \textit{visualizations} of usability problem indicators \cite{fan_vista_2020, soure_coux_2021, batch_uxsense_2023}, (2) \textit{interaction modality} with conversational assistants (CAs) \cite{kuang_collaboration_2023}, (3) \textit{explanations} for problem suggestions \cite{fan_human-ai_2022}, and (4) the \textit{timing} of those suggestions \cite{kuang_enhancing_2024}. 
While these studies provide promising evidence for AI-supported evaluations, they leave open two foundational dimensions that shape real-world collaboration.
The first is the \textbf{time} factor: how collaboration unfolds across repeated sessions as novelty diminishes, familiarity increases, and evaluators develop stable practices. 
Most existing work employs short, single-session studies (35-100 minutes), offering limited insight into how evaluators adapt to AI tools over time. 
Such brief interactions make it difficult to distinguish genuine effects from novelty, as user perceptions and behaviors often shift with continued use \cite{sung_robots_2009, ployhart_longitudinal_2010}.
The second is \textbf{perceived expertise}, operationalized here as the precision and recall of AI-generated suggestions. 
Prior studies typically held the underlying accuracy of AI suggestions constant, whether through Wizard-of-Oz control \cite{fan_human-ai_2022, kuang_collaboration_2023}, by using a general-purpose LLM such as ChatGPT without tailored prompt engineering \cite{kuang_enhancing_2024}, or by employing a custom GPT trained on Nielsen's heuristics \cite{thai_smarter_2025}.
As a result, we know little about how evaluators respond when the AI's expertise varies, even though real-world systems inevitably differ in quality.
Yet effective human-AI collaborative usability analysis depends on a \textbf{joint development of trust and expertise over time}. One can occur without the other, but it is unclear whether meaningful collaboration is possible without both. Because prior work has largely bypassed these dimensions, we still do not know whether human-AI collaborative UX evaluation remains effective when collaboration unfolds over time and when the AI's expertise varies.

To explore the \textbf{time} factor, we examine how collaboration between UX evaluators and CAs unfolds across multiple sessions. 
A longitudinal design allows us to capture how user behavior, trust, and reliance evolve as initial novelty diminishes and familiarity grows, thereby improving ecological validity by accounting for both novelty and learning effects \cite{schmuckler_what_2001}. 
The understudied evolution of human-AI collaborative analysis motivates our first research question (RQ):
\begin{itemize}
\item \textbf{RQ1}: How do UX evaluators' analytic behaviors and attitudes toward a CA change over time?
\end{itemize}
}

\rv{
Meanwhile, advances in prompt engineering show that LLMs can adopt tailored personas that significantly change the accuracy of outputs and shift user interaction dynamics \cite{krapp_quasi-social_2024, shin_can_2025}. 
This raises an important design question for usability analysis, a domain grounded in specialized knowledge of user behavior, heuristics, and design principles \cite{nielsen_heuristic_1994}: how does varying the \textbf{perceived expertise} of the AI system impact evaluators' behavior and attitudes? 
We foreground the \textit{novice-expert framework} as the most theoretically and practically relevant lens for this inquiry. 
Extensive research shows that novices and experts differ systematically in domain knowledge \cite{altexsoft_ux_2019, varnagy-toth_2021_2022}, skills \cite{rynes_experienced_1997, seo_exploration_2021, ironhack_evolving_2023}, and analytic strategies \cite{kuang_merging_2022, kalman_it_2019, foong_novice_2017, schmidt_learner_2020}. 
Because these differences directly shape how evaluators identify and interpret usability problems, expertise provides a meaningful and ecologically grounded design dimension compared to communication style (which is often superficial) or underlying model capability (which is commonly opaque and not tunable in commercial systems).
This focus on expertise is also motivated by \textit{real-world practice}. 
Commercial AI tools for usability analysis vary widely in their apparent expertise due to differences in underlying models (e.g., UserTesting combines proprietary and open-source models \cite{usertesting_usertesting_2020}) and the data they can access (e.g., Maze draws on accumulated transcripts \cite{tanovic_ai_2025}). 
As UX teams increasingly mix tools within their workflows, they will encounter AI systems with inconsistent levels of expertise, making variable AI expertise a practically relevant and emerging design consideration.

Building on this foundation, we investigate how prompting a CA to simulate either a novice or expert UX evaluator shapes human-AI collaboration in usability analysis. 
Prior work suggests that expert-like assistants and high-accuracy AI systems can enhance trust and insight generation \cite{yu_user_2017, persky_moving_2017}, making the design of an experienced CA that provides accurate and comprehensive usability problem descriptions a natural goal. 
However, over-reliance on AI is a well-documented concern in AI-assisted decision-making, as it can reduce critical engagement and impair human performance \cite{vasconcelos_explanations_2023, bucinca_trust_2021}.
Novice-like assistants, by occasionally making mistakes, may counteract this risk by prompting evaluators to reflect more critically on AI suggestions. 
This idea aligns conceptually with the \textit{protégé effect} in education, where teaching others promotes deeper learning \cite{hald_error_2021}, and with the \textit{rubber duck effect} in programming, where explaining code aloud to an object with no knowledge fosters critical reflection and better debugging outcomes \cite{kim_rubber_2020}.
Novices may also bring a fresher perspective, as they are less constrained by established knowledge frames that can anchor experts \cite{foong_novice_2017}.
Since both novice and expert CAs may provide distinct benefits, and little is known about how perceived AI expertise influences evaluators over time, this motivates our second RQ: 
\begin{itemize}
\item \textbf{RQ2}: How does the perceived expertise of a CA (simulated as either a novice or experienced UX evaluator) influence UX evaluators' analytic behaviors and attitudes over time?
\end{itemize}

}

To address these RQs, we developed a usability analysis tool that incorporated two different CAs. 
\rv{Advances in LLMs enable the creation of custom CAs tailored to specific tasks \cite{openai_introducing_2023}, and prior work has applied them to design contexts, including DesignGPT and Idea Generation GPT \cite{ding_designgpt_2023, wang_aideation_2025}.}
Building on this capability, we designed two custom GPTs to simulate novice and experienced UX evaluators, using the novice-expert paradigm as a prompt-generating framework to elicit distinct yet valuable forms of feedback across the two conditions. 
The CAs were integrated into a video analysis tool that supported common usability evaluation tasks: identifying problems, diagnosing causes, suggesting redesigns, and answering user queries. 
To isolate the contribution of CA support, we implemented three experimental conditions: a baseline with no CA, a novice CA, and an experienced CA.
We conducted a within-subjects study with twelve UX evaluators, each completing five analysis sessions over three weeks (approximately six hours of interaction with the tool).
We analyzed their strategies by examining video playback behaviors (e.g., pauses, number of passes) and interactions with the CAs (e.g., accepting or ignoring problem suggestions). 
Analytic performance was measured by the total number of usability problems identified, the number of unique problems found, and inter-rater reliability. 
Finally, we assessed participants' attitudes towards the CAs through Likert scale ratings of perceived efficiency, completeness, trust, confidence, and UX expertise.


For \textbf{RQ1}, we identified two categories of analysis strategies: ``two-pass'' (e.g., reviewing a video twice) and ``one-pass'' (e.g., reviewing once). Two-pass strategies were more common during early sessions, while one-pass strategies became more prevalent over time. 
Participants exhibited a clear \textit{novelty effect}, reporting higher perceived efficiency and trust in the first session, followed by a dip in the second, and a recovery in the third session, which sustained until the end.
For \textbf{RQ2}, participants consistently preferred the experienced CA, rating it significantly higher in efficiency, trustworthiness, and the completeness of its suggestions. 
In contrast, the novice CA was seen as more useful for training, prompting participants to critically engage with its suggestions rather than accepting them at face value. 
Importantly, participants were unable to distinguish between the perceived expertise levels of the two CAs until they had more hands-on experience in the second session.
Our findings across both RQs highlight the value of longitudinal studies in human-AI collaboration and point to different use cases for novice and experienced CAs. 
Furthermore, we provide implications for future longitudinal studies on human-AI collaboration. 
In summary, our contributions include:
\begin{itemize}
    \item Development of an LLM-based analysis tool with novice and experienced CAs designed to identify usability problems and respond to user questions;
    \item Findings from a multi-session study illustrating how UX evaluators' analytic behaviors and perceptions during human-AI collaborative analysis progress over time;
    \item Comparative analysis of novice and experienced CAs, examining their differing impacts on usability analysis.
\end{itemize}
\section{Related Work}

Our work draws inspiration from related research in two key areas: 1) longitudinal studies on human-AI collaboration, and 2) collaboration with CAs featuring varying levels of expertise.

\subsection{Longitudinal Studies on Human-AI Collaboration}

Longitudinal studies involve gathering data over multiple time points, incorporating time as a dependent variable into the research design \cite{gerken_longitudinal_2010}. 
In HCI, these studies deepen our understanding of how human-technology interactions evolve over time \cite{kjaerup_longitudinal_2021}.

\subsubsection{Benefits of Longitudinal Studies} 

Longitudinal studies address the \textit{novelty effect}, where users' initial responses to technology differ from long-term usage patterns \cite{ployhart_longitudinal_2010}. 
Previous research shows that initial fascination with new technologies often diminishes, leading to discontinued use once the novelty effect subsides \cite{kanda_two-month_2007, tanaka_daily_2006, sung_robots_2009}. 
At the same time, extended use can also foster positive outcomes \cite{mutsuddi_text_2012, shin_beyond_2019, vertsberger_adolescents_2022}. 
For example, over four months, older adults' perceptions of a smart speaker evolved from appreciating its simplicity to valuing it as a digital companion, showing a shift from superficial interaction to meaningful engagement \cite{kim_exploring_2021}. 
Their initial challenges with the device diminished as they adapted their troubleshooting behaviors, like repeating or paraphrasing commands, instead of blaming the device \cite{kim_exploring_2021}. 
These findings highlight how user behaviors and perceptions continue to evolve after the novelty effect fades, with adoption trajectories varying widely across contexts. 
In our study, we therefore focused on the usability context and how evaluators adapted their strategies over time as they developed a better understanding of the CAs' capabilities.

\subsubsection{Research Gap in Longitudinal Studies on Human-AI Collaboration}

Recent advances in AI have spurred the development of human-AI collaboration tools \cite{luger_like_2016}. 
However, a review of 37 empirical studies on human-AI collaboration from 2018 to 2023 reveals a heavy reliance on short-term, quantitative experiments, with relatively few studies employing longitudinal or qualitative methods \cite{pentina_consumermachine_2023}. 
This lack of methodological diversity may be due to the novelty of AI technologies and the practical challenges of conducting long-term research, such as participant retention and time investment \cite{gerken_longitudinal_2010, karapanos_theories_2012}.
Yet relying solely on short-term experiments can produce misleading conclusions. 
These studies typically capture participants' initial reactions to AI tools, which may be shaped by curiosity and unfamiliarity, rather than sustained, stable patterns of use \cite{sung_robots_2009}. 
For instance, early interactions with CAs often evoke feelings of social connection or anthropomorphism \cite{ho_psychological_2018}, but it remains unclear whether these impressions endure with repeated exposure \cite{araujo_speaking_2024}. 
Some longitudinal studies of social or health-focused CAs have shown diminishing engagement over time \cite{croes_can_2021}, while others have found increasing trust with continued use \cite{araujo_speaking_2024}. However, these findings have not been extended to task-oriented domains like usability evaluation, where sustained performance and domain expertise are crucial.

To address this gap, our study takes a longitudinal approach to examine how evaluators' interactions with a CA evolve over time. 
Unlike short, one-off evaluations, multi-session studies enable us to observe shifts in trust, reliance, and analytic strategies that emerge through repeated use. 
For example, trust in AI is known to accumulate gradually, influenced by past experiences and evolving expectations \cite{kahr_understanding_2024}. 
By studying UX evaluators across five sessions over three weeks, we capture both their immediate impressions and how they adapt their evaluative practices, respond to AI suggestions, and form sustained working relationships with CAs. 
\rv{This approach provides an ecologically valid and comprehensive view of how human-AI collaboration unfolds in practice.}

\subsection{Collaborating with Novice and Experienced CAs}
\label{sec:experience-differences}

\subsubsection{Existing Human-AI Collaborative Tools for Usability Analysis}
In exploring the role of AI as an \textit{assistant} to UX evaluators, prior research has employed two approaches: representing AI as non-interactive visualizations \cite{fan_vista_2020, fan_human-ai_2022, soure_coux_2021, batch_uxsense_2023}, and developing interactive conversational assistants (CAs) \cite{kuang_collaboration_2023, kuang_enhancing_2024, kuang_evaluating_2025}. 
Visualizations, such as icons and line charts, effectively highlight usability problems like abnormal pitch and negative sentiment \cite{fan_automatic_2020}. 
However, these static representations limit UX evaluators' ability to engage with the data and ask follow-up questions for AI-generated results.
CAs address some of these limitations by enhancing interactivity; they offer usability problem suggestions and provide explanations on demand \cite{kuang_collaboration_2023, kuang_enhancing_2024}. 
Research has shown that UX evaluators prefer text-based interactions over voice when analyzing usability videos \cite{kuang_collaboration_2023}, and that proactive CAsthat automatically suggest usability problems at specific timestamps can further support evaluators \cite{kuang_enhancing_2024}. 

However, while these advancements demonstrate the benefits of both visualization and interactivity, there remains an unaddressed gap: previous systems have not accounted for \textit{variations in AI's expertise} \cite{fan_human-ai_2022, soure_coux_2021, kuang_enhancing_2024, shen_multiux_2026}. 
This omission raises concerns about the generalizability of the findings, as participants' reactions are likely influenced by the AI's perceived UX expertise. 
For instance, a recent study demonstrated that both trust and reliance on AI advice were significantly higher when the model had high accuracy compared to low accuracy \cite{kahr_understanding_2024}. 
\rv{Prior studies on AI-assisted usability analysis have largely relied on generic versions of ChatGPT without prompt customization. For example, some researchers pasted usability test transcripts directly into ChatGPT and asked it to identify usability problems \cite{kuang_enhancing_2024}, while others used GPT-4 to automate heuristic evaluations and detect UI violations without defining a specific role or expertise for the model \cite{duan_generating_2024}. 
Although these approaches demonstrated the potential of conversational AI to support analysis, they also revealed the limitations of using general-purpose models without UX-specific expertise.
In contrast, a recent short paper systematically compared five prompting techniques: zero-shot prompting, role prompting, chain-of-thought prompting, self-refine prompting, and least-to-most prompting, for identifying usability problems on a shopping website. 
The results showed that role prompting was particularly effective in uncovering problems, providing contextual analyses, and generating actionable insights from a user perspective \cite{shin_can_2025}. 
Building on these insights, our work investigates how perceived AI expertise (operationalized through role prompting) influences users' engagement with and utilization of the system over time.}


\subsubsection{Motivation for Simulating CAs as Novice and Experienced UX Evaluators}

The \textit{novice-expert paradigm} explains that novices learn and think differently from experts \cite{lieberei_findings_2023}, progressing through five stages of proficiency: novice, advanced beginner, competent, proficient, and expert \cite{benner_novice_1982}. 
Differences between novices and experts have been observed in various fields, including product and UI design \cite{chen_behaviors_2022, cockburn_supporting_2014}, pair programming \cite{lui_pair_2006}, and aerospace engineering \cite{deken_tapping_2011}. 
In the context of UX, expertise is defined as the ability to apply knowledge effectively in UX work, with previous research categorizing UX professionals as either ``early career'' or ``later career'' \cite{seo_exploration_2021}. 
Given the varying terminologies used to describe levels of expertise \cite{fisher_defining_1991}, we adopt the general terms ``\textbf{novice}'' and ``\textbf{experienced}'' in this paper.

A key motivation for simulating CAs as novice and experienced UX evaluators is the significant differences in how UX evaluators approach usability analysis based on their level of expertise \cite{kuang_merging_2022}. 
Experienced UX evaluators possess better-organized knowledge and more effective strategies for accessing and applying that knowledge, enabling them to select and implement the most appropriate research methods \cite{persky_moving_2017, rynes_experienced_1997}. 
In contrast, novice evaluators often have a narrower range of techniques and a more limited understanding \cite{altexsoft_ux_2019}. 
However, they may also bring up-to-date training that is valuable for navigating ongoing digital transformation \cite{ironhack_evolving_2023}.
\rv{Furthermore, novices are less anchored to established ``frames'' (i.e., cognitive structures used to interpret events or data), allowing them to offer a fresher perspective \cite{foong_novice_2017}.
This aligns with the ``curse of knowledge,'' where experts may assume others have the same knowledge, introducing bias and constraining their viewpoints \cite{tullis_curse_2022, xiong_curse_2020}.}
When analyzing user feedback, experts are more likely than novices to question inconsistencies and seek critical information to deepen their understanding of design goals \cite{foong_novice_2017}. 
Additionally, experienced UX evaluators often customize usability problem descriptions and collaborate to ensure reliability \cite{folstad_analysis_2010, kuang_merging_2022}, whereas novice researchers may struggle with data analysis, such as identifying meaningful patterns from codes \cite{kalman_it_2019} and analyzing complex problems \cite{schmidt_learner_2020}. 

While designing an experienced CA that provides accurate and complete usability problem descriptions is a clear goal, over-reliance on AI remains a well-documented concern in AI-assisted decision-making, as it can impair human performance and reduce critical engagement \cite{vasconcelos_explanations_2023, bucinca_trust_2021}. 
Various strategies have been proposed to mitigate over-reliance and increase human autonomy, such as foregrounding choices and prompting reflection \cite{clasen_fostering_2024} or introducing cognitive forcing functions to encourage independent judgment \cite{bucinca_trust_2021}. 
In our study, we explore an alternative approach: simulating a novice CA that identifies fewer usability problems compared to the experienced CA.
This design encourages users to stay alert, critically evaluate the AI's suggestions, and take a more active role in the analysis process \cite{hald_error_2021}.
Rather than intentionally weakening the AI, the novice CA is framed as a collaborative peer-in-training. 
This framing draws inspiration from the \textit{protégé effect}, a phenomenon in which individuals tend to think more deeply and critically when they are responsible for teaching, supporting, or correcting someone else's understanding \cite{chase_teachable_2009}. 
While the protégé effect has been primarily studied in educational settings, its core mechanism---heightened cognitive engagement driven by a sense of responsibility---may extend to collaborative work, including human-AI interaction. 
For instance, studies have found that when individuals felt accountable for teaching others, even fictional characters, they were more likely to engage more critically with the material and acknowledge errors \cite{chase_teachable_2009, corwin_teaching_2023}. 
\rv{In our context, a novice CA may elicit a mentoring mindset in UX evaluators, leading them to examine the AI’s suggestions more critically, reflect more deeply on the analysis, and develop a more nuanced understanding of usability problems. 
We investigate this as a possible mechanism through which novice CAs could offer distinct benefits.}
\section{Design of the Conversational Assistants}

To answer our RQs on the long-term use of CAs in usability analysis and the impact of perceived UX expertise, we first needed to simulate a novice CA and an experienced CA. 
Building upon prior work on CAs for usability analysis, we designed two types of support:
\begin{enumerate}
    \item \textbf{Automatic suggestions} that appear \textit{after} a usability problem occurred in the video. Findings in prior work indicated that presenting suggestions after potential usability problems significantly enhanced perceived efficiency, trust, and user preference \cite{kuang_enhancing_2024}. Recognizing the participants' feedback for the CA to not only describe the problem but also identify its root cause, we have included information on the \textit{cause of the problem}. Additionally, each identified problem is accompanied by a \textit{redesign recommendation}, aligning with the overarching goal of usability testing to unearth opportunities for design improvement \cite{moran_usability_2019}. This inclusion of causes and redesign recommendations distinguishes our CAs from previous work \cite{kuang_collaboration_2023, kuang_enhancing_2024}. 
    \item \textbf{Reactive responses} to UX evaluators' questions. Building on prior research showing that UX evaluators seek to ask questions during analysis, and using a dataset of common queries \cite{kuang_collaboration_2023}, the CA is designed to provide answers to various questions posed by evaluators about usability videos.
\end{enumerate}

The following subsections describe the process of simulating the CAs, generating automatic suggestions and reactive responses, and evaluating our approach. 

\subsection{Simulating Novice and Experienced CAs}
We adopted the novice-expert paradigm as a prompt-engineering framework, informed by our literature review on differences in UX knowledge and analysis practices between novice and experienced evaluators (Section~\ref{sec:experience-differences}). 
Building on recent work showing that LLMs can simulate diverse user characteristics \cite{shin_understanding_2024, xiang_simuser_2024, krapp_quasi-social_2024, lauer_vector_2024} and be prompted for specific design tasks (e.g., DesignGPT, Idea Generation GPT, Keyword Extraction GPT, \rv{Nielsen-heuristics-trained GPT}) \cite{ding_designgpt_2023, wang_aideation_2025, thai_smarter_2025}, we created two custom GPTs to simulate novice and experienced UX evaluators, both powered by \rv{GPT-4V} (see Table~\ref{tab-chp7:CA-characteristics}). Task instructions were defined using OpenAI's GPT Builder interface on the ChatGPT web platform, following the ``Creating a GPT'' guidelines \cite{openai_creating_2024} (see Appendix, Table~\ref{tab-chp7:GPT-prompts}).
Prior studies have shown that LLMs can approximate domain experts in contexts such as research ideation and games, producing overlapping input across expertise groups \cite{tyni_can_2024, liu_personaflow_2025}. 
In contrast, our aim was not to test how faithfully GPT could replicate novice or experienced UX evaluators. 
Rather, we sought to generate two CAs with clearly distinguishable capabilities in usability analysis, ensuring a meaningful contrast in \textit{perceived expertise}. 
To contextualize the effects of this tailoring, we also included the general version of \rv{GPT-4V\footnote{GPT-4 with vision was the most advanced model available at the time of the study.}} (without prompt customization) as a comparison point for evaluating the added value of expertise design.

\begin{table}[h]
  \centering
  \caption{Prompts Used to Simulate Novice and Experienced CAs}
  \label{tab-chp7:CA-characteristics}
  \small
  \begin{tabular}{p{1.25cm} p{6.2cm}}
    \toprule
    \textbf{Condition} & \textbf{Prompt [with references to relevant sources]} \\
    \midrule
    Novice & 
    You are a novice UX evaluator with limited hands-on experience in conducting user research and a basic understanding of UX research methodologies and concepts \cite{altexsoft_ux_2019}. Although you are trained in the latest tools \cite{ironhack_evolving_2023}, you may find it challenging to analyze data, such as identifying meaningful patterns from codes \cite{kalman_it_2019} and tackling complex problems \cite{schmidt_learner_2020}. \\ 
    \midrule
    Experienced & 
    You are an experienced UX evaluator with extensive hands-on experience in conducting user research and a deep understanding of UX research methodologies and concepts \cite{rynes_experienced_1997}. You frequently customize usability problem descriptions and prioritize ensuring reliability in your analyses \cite{folstad_analysis_2010, kuang_merging_2022}. Additionally, you actively question inconsistencies and seek critical information to deepen your understanding of design goals \cite{foong_novice_2017}. \\ 
    \bottomrule
\end{tabular}
\end{table}

\subsection{Generating Automatic Usability Problem Suggestions}

We compiled a dataset comprising 15 usability videos covering three products, each having five users completing the same tasks. Details of the study videos are in Section \ref{sec:study-videos}. 
Initially, transcripts for these videos were automatically generated by Zoom, as the recordings were made using this software. 
Subsequently, we thoroughly reviewed the transcripts to correct any errors, add punctuation, and adjust timestamps to reflect natural speech breaks.

Following recent work that demonstrated GPT's ability to understand UI page information \cite{liu_make_2024}, we first uploaded screenshots of the main pages of the product directly to the GPT webpage\footnote{\href{https://chatgpt.com/gpts}{https://chatgpt.com/gpts}} so that the custom GPTs could develop an understanding of the UI elements available for users to interact with. 
For example, we uploaded a screenshot of the home page of a test website (Fig. \ref{fig-chp7:UI-understanding}-A in the Appendix). 
The ``experienced GPT'' provided a response that described the content of the screenshot, including a title bar, input field, example posters, poster thumbnails, footer link, preview pane, user profile icon, and closing option (Fig. \ref{fig-chp7:UI-understanding}-B in the Appendix). 
The detailed description shows that the GPTs could extract relevant information from the provided screenshots. 
We then provided the GPTs with the tasks completed by each user in the video, along with the transcript, and prompted them to generate usability problem descriptions, causes, and redesign recommendations in a prescribed format. 
Table \ref{tab-chp7:GPT-prompts} in the Appendix shows the specific prompts, which followed OpenAI's official prompt engineering guide (e.g., including relevant details in the query, asking the model to adopt a persona, and using delimiters to clearly separate distinct parts of the input) \cite{openai_prompt_2024}. 
Prior work suggests that generating responses three times can improve the stability of ChatGPT's output performance \cite{giannakopoulos_evaluation_2023, ronanki_chatgpt_2024}. 
Thus, we prompted the GPTs three times for each video and used the union of all problems generated across these three rounds as the final list.
To generate usability problem suggestions for the general version, we used the same process as for the custom GPTs, but directly through the ChatGPT interface without any prompt customization.

\subsection{Generating Responses using the OpenAI API}

To meet user expectations and in line with the behavior of common voice assistants like Siri and Google Assistant \cite{luger_like_2016}, the novice and experienced CAs were designed to respond to impromptu questions posed by UX evaluators.
In contrast to prior work that utilized the Wizard of Oz method \cite{kuang_collaboration_2023, kuang_enhancing_2024}, we leveraged the OpenAI API (using the ``\rv{gpt-4-0613}'' model) to provide real-time responses during each study session \cite{openai_openai_2024}. 
To maintain consistency between real-time responses and pre-generated automatic suggestions, the same prompts used for simulating the respective CA were provided at the beginning of each session (Table \ref{tab-chp7:CA-characteristics}). 
We also supplied the video background, transcript, and the pre-generated list of automatic suggestions so the CA understands the video content (Table \ref{tab-chp7:GPT-prompts}).
For the general version, we directly provided the video transcript and suggestions list without prompting to adopt a persona.

\subsection{Evaluation of Suggestions and Responses}
\label{sec:performance-evaluation}

To assess differences in usability analysis between the two custom CAs and the general version, three authors with UX expertise independently analyzed the study videos. They then held a group discussion to consolidate their findings and resolve any disagreements, following established UX practices \cite{hertzum_evaluator_1998}. 
The finalized list of usability problems from this manual analysis served as the \textit{ground truth} for evaluating the performance of both the custom GPT models and the OpenAI API responses. 
This method of comparing AI-generated results to manual analysis is consistent with prior research on AI-powered usability analysis tools \cite{kuang_enhancing_2024}.
For automatic suggestions, \rv{we manually evaluated whether each suggestion matched a usability problem in the ground truth. Based on this assessment, } 
we calculated precision (the proportion of correct problems among all identified problems) and recall (the proportion of correct problems among all correct problems) for each video. 
For responses, we asked one example question from each of the five main categories previously used by UX evaluators when interacting with CAs \cite{kuang_collaboration_2023}. 
\rv{We then rewatched the corresponding video segments to determine whether the response was correct and} calculated the accuracy as the proportion of correct responses per video. 
The mean and standard deviation across 15 videos are presented in Table \ref{tab-chp7:suggestion-evaluation}. 

\begin{table}[h]
  \centering
  \caption{Suggestion precision and recall, and response accuracy for the three conditions, reported in ``Mean (Standard Deviation)''}
  \label{tab-chp7:suggestion-evaluation}
  \small
  \begin{tabular}{p{2cm} p{1.7cm} p{1.7cm} p{1.7cm}}
    \toprule
    \textbf{Condition} & \textbf{Suggestion Precision} & \textbf{Suggestion Recall} & \textbf{Response \newline Accuracy} \\
    \midrule
    Novice & 0.80 (0.13) & 0.54 (0.13) & 0.73 (0.16) \\ 
    \midrule
    Experienced & 0.93 (0.08) & 0.76 (0.09) & 0.88 (0.17) \\ 
    \midrule
    General & 0.84 (0.12) & 0.59 (0.13) & 0.76 (0.19) \\
    \bottomrule
\end{tabular}
\end{table}

We found that the precision of the novice CA was 0.80, while the general version had a precision of 0.84, meaning 20\% and 16\% of the identified problems were false positives, respectively. 
In contrast, the experienced CA had a precision of 93\%, with only 7\% being false positives. 
All conditions had lower recall than precision, with the novice CA identifying just 54\% of all ground truth problems and the experienced CA identifying 76\%.
This pattern is comparable to the evaluation of ChatGPT's performance in prior work \cite{kuang_enhancing_2024}, where precision was higher than recall, suggesting that while not every problem was detected, the ones identified were relevant.

To compare the three conditions, we conducted a one-way ANOVA on suggestion precision ($F_{2, 28} = 17.3, p < .0001, \eta_{p}^{2} = 0.55$), recall ($F_{2, 28} = 82.3, p < .0001, \eta_{p}^{2} = 0.85$), and response accuracy ($F_{2, 28} = 4.0, p < .05, \eta_{p}^{2} = 0.22$), all of which showed significant differences. 
Posthoc pairwise comparisons revealed significant differences between the novice and experienced CAs in suggestion precision ($p < .01$), recall ($p < .0001$), and response accuracy ($p < .05$).
However, the differences between the general version and the novice CA were not significant (all $p > .05$). 
Significant differences were only found between the general version and the experienced CA in precision ($p < .05$) and recall ($p < .001$), but not in response accuracy.
Thus, the experienced CA achieved higher precision, recall, and response accuracy compared to both the novice and general versions, which performed similarly.

\begin{table*}[h]
  \centering
  \caption{Example usability problem suggestion for the three conditions, with differences bolded and underlined}
  \label{tab-chp7:content-evaluation}
  \small
  \begin{tabular}{p{1.5cm} p{3.5cm} p{4cm} p{4cm}}
    \toprule
    \textbf{Condition} & \textbf{Problem Description} & \textbf{Cause of Problem} & \textbf{Redesign Recommendation} \\
    \midrule
    Novice & 
    Confusion about how to save the current design. & 
    The platform did not provide clear visual feedback or confirmation when a design was saved. & 
    Include a `Save' button and visual confirmation that the design has been saved. \\ 
    \midrule
    Experienced & 
    The user is unsure if the design is automatically saved \textbf{and how to proceed with caption generation}. & 
    There were no clear visual cues or confirmations indicating that work is saved, \textbf{nor a clear transition to the next steps for social media sharing.} & 
    \ul{Introduce an auto-save feature} and corresponding notifications, \textbf{and a clear `Next Steps' feature that leads users to caption and hashtag generation after the poster is saved.} \\ 
    \midrule
    General & 
    The user is unsure if their work is being automatically saved. & 
    Lack of clear visual feedback or a message indicating that the design was saved. & 
    \ul{Introduce an auto-save feature} with a clear indicator and offer a manual save option. \\
    \bottomrule
\end{tabular}
\end{table*}

We also compared the suggestion content across the three conditions when they identified the same problem. 
\rv{To minimize confounding factors such as wording, semantics, or tone, our prompts instructed the CAs to provide responses in a consistent format, without assigning any personality or stylistic variations (Table \ref{tab-chp7:GPT-prompts}). 
Table \ref{tab-chp7:content-evaluation} shows representative examples of suggestions from each condition, demonstrating that the structure and tone of the outputs were comparable.}
In the example from Table \ref{tab-chp7:content-evaluation}, the experienced CA identified an additional issue (e.g., how to proceed with caption generation) and provided more specific causes and redesign recommendations. 
The novice and general versions were largely similar, although the general version recommended an auto-save feature in addition to the manual save option.
Since the novice CA performed comparably to the general version in terms of precision, recall, accuracy, and content quality, we excluded the general version from the user study. 
We focused on the novice and experienced CAs, which aligns with Nelson and Stolterman's concept of the ``ultimate particular,'' which emphasizes engaging with specific, well-defined design instances rather than generic or universal ones \cite{nelson_ultimate_2012}. 
In our case, the novice CA served as a ``floor'' and the experienced CA as a ``ceiling,'' creating the strongest contrast in perceived expertise. This contrast allowed us to examine how evaluators' analytic performance and perceptions evolved across distinct levels of CA capability. 
Furthermore, limiting the study to two CA conditions helped reduce participant workload and session length, thereby lowering the risk of dropout.
\section{Usability Analysis Tool}

\begin{figure*}[h]
  \centering
  \resizebox{\textwidth}{!}{%
        \scalebox{1}{\includegraphics{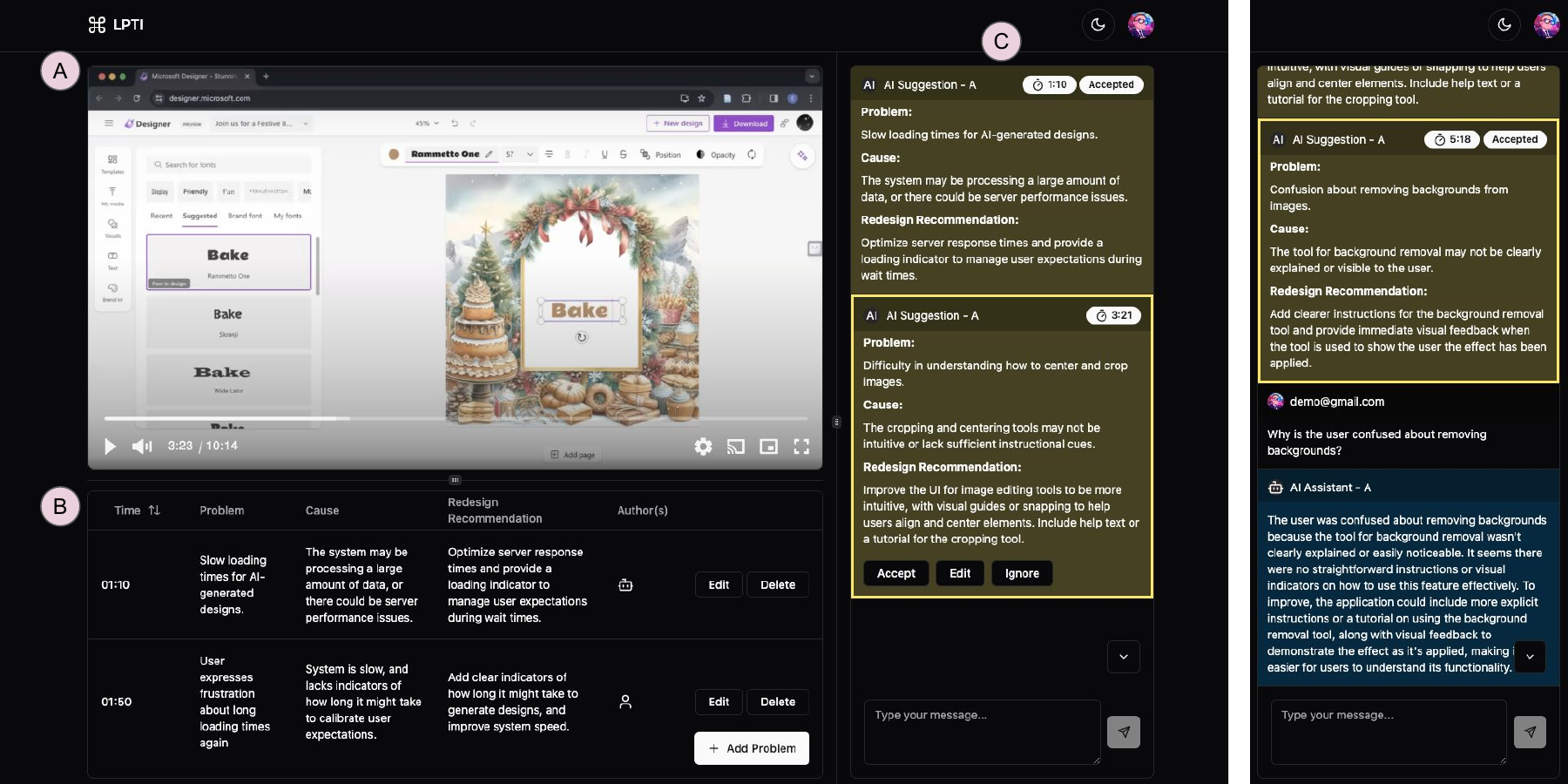}}
    }
  \caption{User interface of the usability analysis tool for the conditions with CAs containing: A) video player, B) table of usability problem descriptions, causes, and redesign recommendations with the corresponding timestamp, and C) chat thread.}
  \Description{Screenshot of the user interface. Section A displays a video player paused at 3:23. Section B is a table listing problems, causes, redesign recommendations, and author details for different problems. Section C features a chat thread with the options to accept, edit, or ignore an AI suggestion, a badge that indicates the suggestion's status, the suggestion's timestamp, a text input field, and the CA's response.}
  \label{fig-chp7:prototype-UI}
\end{figure*}

Once we had created the two CAs, we needed to build a tool that allowed UX evaluators to interact with the CAs while analyzing usability videos. 
In line with prior human-AI collaborative analysis tools \cite{kuang_collaboration_2023, kuang_enhancing_2024}, the usability analysis tool includes a \textbf{video player} for UX evaluators to review recordings (Fig. \ref{fig-chp7:prototype-UI}-A). 
Additionally, we introduced a \textbf{problem table} (Fig. \ref{fig-chp7:prototype-UI}-B) to facilitate easier management of problem descriptions by UX evaluators. 
In conditions where participants use a CA, a \textbf{chat window} displays automatic suggestions and the conversation thread (Fig. \ref{fig-chp7:prototype-UI}-C). 
The baseline condition lacks this chat window (Fig. \ref{fig-chp7:prototype-UI-baseline} in the Appendix). 
The chat window is used exclusively for communicating with the CA, while the problem table serves to collect confirmed usability problems identified by UX evaluators.

When automatic suggestions appear for the first time, they come with three action buttons: 
\begin{enumerate}
    \item \textbf{Accept}, which directly adds the suggestion to the problem table
    \item \textbf{Edit}, which opens a modal box for UX evaluators to modify the suggestion before adding it to the problem table
    \item \textbf{Ignore}, which does not add anything to the problem table
\end{enumerate}
The design of these actions is based on research on AI-assisted decision-making tools, which outlines typical responses to AI recommendations: accept/trust, ignore/reject, or edit/negotiate \cite{sivaraman_ignore_2023}. 
Once the UX evaluator selects an action button, the suggestion is updated with a status badge (e.g., \includegraphics[height=0.3cm]{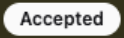}). 
The corresponding usability problem then appears in the Problem Table with a robot icon, indicating it was authored by the CA (\includegraphics[height=0.3cm]{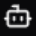}).

In addition to adding problems based on automatic suggestions, UX evaluators can click the ``\textbf{+ Add Problem}'' button (\includegraphics[height=0.3cm]{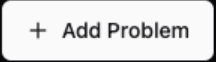}). 
This action opens an empty modal form (Fig. \ref{fig-chp7:prototype-UI-baseline}), allowing UX evaluators to manually enter usability problem descriptions, causes, and redesign recommendations. 
These manually added problems then appear in the Problem Table with a user icon, indicating manual addition (\includegraphics[height=0.3cm]{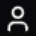}). 
For problems edited by UX evaluators, both robot and user icons are displayed side by side.

When UX evaluators click on the timestamp at the top right corner of each suggestion (e.g., \includegraphics[height=0.3cm]{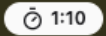}), the video player jumps to the corresponding timestamp, allowing UX evaluators to easily navigate and rewatch specific video segments. 
Additionally, if UX evaluators have questions about a suggestion, they can type their messages into the text input field and receive responses. 
Responses appear in blue to distinguish them from automatic suggestions, which are shown in yellow.
Finally, our interface supports both light and dark modes (Fig. \ref{fig-chp7:prototype-UI-light} in the Appendix).
\section{User Study}

To answer our RQs on the long-term use of CAs in usability analysis and the impact of different UX experience levels, we conducted an IRB-approved within-subjects study with three conditions: 1) no CA (baseline), 2) novice CA, and 3) experienced CA. 

\subsection{Participants and Apparatus}
\begin{figure*}[h]
  \centering
  \resizebox{\textwidth}{!}{%
        \scalebox{1}{\includegraphics{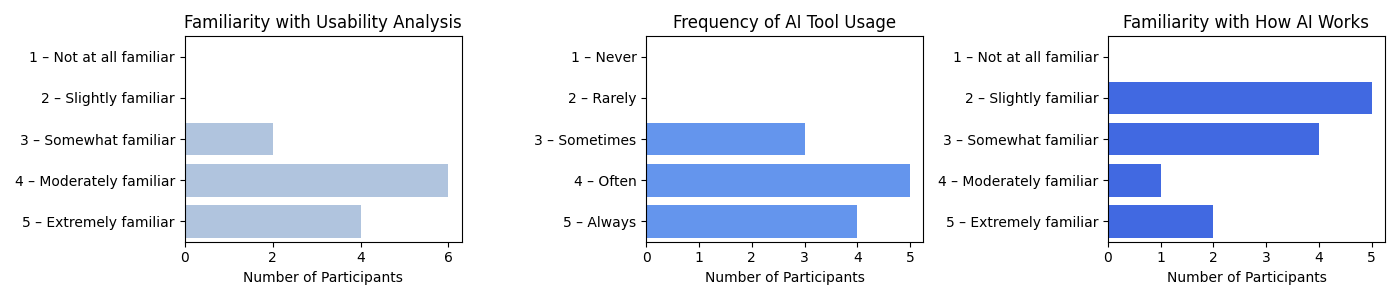}}
    }
  \caption{Bar charts showing participants' familiarity with usability analysis, frequency of AI tool usage, and familiarity with how AI works.}
  \Description{Three horizontal bar charts showing participants' familiarity with usability analysis (ranging from 3 - somewhat familiar to 5 - extremely familiar), frequency of AI tool usage (ranging from 3 - sometimes to 5 - always), and familiarity with how AI works (ranging from 2 - slightly familiar to 5 - extremely familiar).}
  \label{fig-chp7:demographics}
\end{figure*}

We recruited 12 (10 females and 2 males) participants with an average of 4.6 years of UX experience (SD = 2.1). 
\rv{Eight participants were UX professionals, and four were HCI graduate students who had completed UX internships.}
Their average age was 28 years (SD = 4). Figure \ref{fig-chp7:demographics} shows participants' familiarity with usability analysis (Md = 4, IQR = 0.5), frequency of AI tool usage (Md = 4, IQR = 1.5), and understanding of how AI works, including how LLMs generate responses (Md = 3, IQR = 1.5).
All participants had prior experience using AI tools, \rv{such as  ChatGPT, for tasks like content generation, writing, and qualitative analysis (e.g., surveys, interviews)}.
However, most lacked a deep understanding of how these tools work.
The study sessions were conducted remotely through Zoom, and video recordings were captured for subsequent analysis. 

\subsection{Study Videos}
\label{sec:study-videos}

Since there is no standardized taxonomy of products for evaluating usability analysis tools, we curated a set of usability videos to facilitate analysis, following previous research (e.g., \cite{fan_vista_2020, fan_human-ai_2022, soure_coux_2021, kuang_collaboration_2023}). 
To ensure diversity in interface types, we selected three distinct platforms---a desktop website, a smartphone app, and a VR headset---with five users testing each product, yielding a total of 15 videos for analysis.
The rationale behind involving five users for testing each product aligns with the optimal sample size for usability studies, grounded in a cost-benefit analysis \cite{nielsen_why_2000, martin_why_2016}.

The tasks in the video focused on the fundamental functionalities of the respective products, such as seeking the weather forecast on a weather app or creating a poster using a generative AI (GenAI) design tool.
The users in the videos were either recruited from the community or were students from the authors' institution.
As they had no prior experience with these products, they encountered a wide range of usability issues, providing rich material for evaluating usability analysis approaches. 
Table \ref{tab-chp7:videoInfo} in the Appendix details each video, including its duration, associated tasks, and the number of problem suggestions from each CA. 
The novice CA found an average of 6.1 usability problems ($SD=1.2$) per video, while the experienced CA found an average of 7.7 problems ($SD=1.8$), which aligned with our evaluation in Section \ref{sec:performance-evaluation}.
Paired Bonferroni-corrected t-tests revealed significant differences between the two conditions ($p < .01$).
By ensuring that each video contained at least 4 usability problem suggestions, we aimed to provide participants with ample material for analysis and interactions with the CA.

\subsection{Procedure}
\label{sec:procedure}

\rv{The study used a within-subject design in which all 12 participants completed five sessions, each reviewing three videos, one per condition. 
Across five sessions, each participant analyzed 15 unique videos (5 sessions $\times$ 3 videos/session = 15 videos) and spent over 6 hours with the tool.
The three conditions (no CA, novice CA, and experienced CA) were counterbalanced using a balanced Latin square design \cite{schwind_hci_2023}. 
The videos were similarly counterbalanced to ensure each appeared with different products across the conditions. 
With 15 videos and three conditions, there were 45 unique video-condition combinations (15 $\times$ 3 = 45). 
Thus, each video-condition combination was reviewed by four participants (12 participants $\times$ 15 videos $\div$ 45 video-condition pairs = 4).
}

We did not disclose the differences between the two CA versions to participants, only referring to them as Version A or B.
To ensure this, we reviewed all conversation logs between the participants and the two CA versions to verify that their roles were not inadvertently revealed during the study sessions.
\rv{}

\begin{figure*}[h]
  \centering
  \resizebox{0.8\textwidth}{!}{%
        \scalebox{1}{\includegraphics{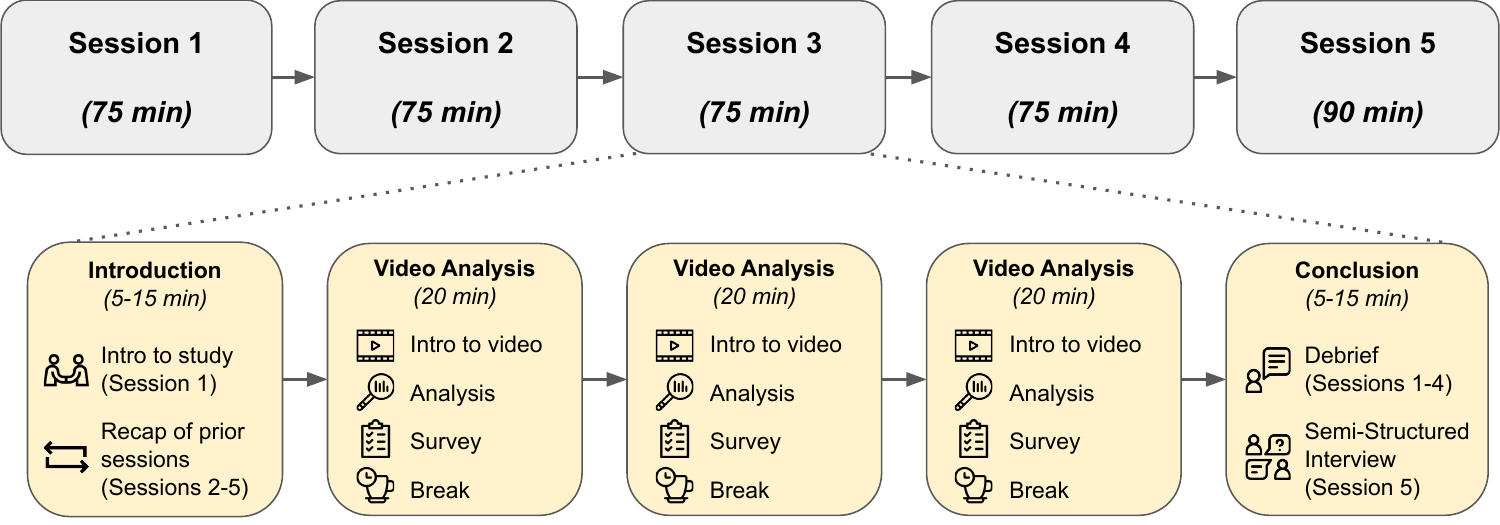}}
    }
  \caption{Flowchart of the longitudinal study containing five sessions in total. Each session included an introduction or recap, an analysis of three videos, and a debrief or semi-structured interview.}
  \Description{Flowchart of the longitudinal study containing five sessions. In each session, there is an introduction (session 1) or recap of prior sessions (sessions 2-4), an analysis of three different videos, and a debrief (sessions 1 - 4) and or semi-structured interview (session 5).}
  \label{fig-chp7:procedure}
\end{figure*}

Figure \ref{fig-chp7:procedure} outlines the study procedure. The first session included an introduction to the study, a tutorial on the tool, and a Q\&A session to address any questions.  
In each session, participants received background information on each of the three videos before starting their analysis, followed by a survey. 
The survey assessed efficiency, trust, suggestion completeness, confidence in analysis, and perceived UX expertise of the CA. 
Sessions 1-4 concluded with a brief debrief to gather feedback and confirm the next session. 
In session 5, we conducted a semi-structured interview to gather participants' overall experience, feedback, and suggestions for improvement.




\subsection{Data Analysis}

We collected the following data: (1) usability problem descriptions, causes, and redesign recommendations; (2) conversation logs between participants and the CAs; (3) interaction logs in the analysis tool; and (4) video recordings of each session.
To analyze usability problem descriptions, we labeled and identified unique usability problems. 
We calculated Fleiss' kappa, a generalization of Cohen's kappa, to assess inter-rater reliability among multiple raters \cite{mendoza_usability_2005}. 
We coded the conversation logs to categorize messages by content. 
We graphed the interaction logs to identify video playback behaviors and subsequent actions after suggestions. 
Semi-structured interview responses were transcribed using speech-to-text software and manually corrected. 
We applied inductive coding to analyze transcripts individually, followed by thematic grouping through discussions \cite{charmaz_constructing_2006}.

For all quantitative data, we used the Shapiro-Wilk test to check the normality. 
To examine the impacts of long-term usage (RQ1) and CA expertise (RQ2), we conducted a two-way ANOVA with two within-subject factors to determine significant differences across \textit{sessions} and \textit{conditions}. 
Effect sizes are reported using partial eta squared ($\eta_{p}^{2}$), and post-hoc pairwise comparisons were made with Bonferroni correction. 
Statistical test results are summarized in Table \ref{tab-chp7:stat-tests} in the Appendix to avoid cluttering the main text.

\section{Results}

The aims of our data analysis are twofold: to understand how UX evaluators' analytic behaviors and attitudes toward a CA evolve (RQ1) and to explore the impact of novice and experienced CAs on UX evaluators' behaviors and attitudes (RQ2). 
We divided the results into three subsections: \ref{sec:analysis-behaviors}: usability video analysis behaviors, \ref{sec:analytic-performance}: analytic performance, and \ref{sec:subjective-feedback}: subjective feedback on human-AI collaborative analysis, with a summary at the beginning of each. 

\subsection{Usability Video Analysis Behaviors}
\label{sec:analysis-behaviors}

\fbox{
  \parbox{0.97\linewidth}{
    \textit{\textbf{Summary for RQ1 (Change over time)}: Participants' analysis strategies evolved from a two-pass to a one-pass approach, enhancing efficiency. A novelty effect emerged, as participants increasingly ignored problem suggestions from the CA after the first session. Moreover, their messages progressed from simple questions to more complex task requests.}
    
    \textit{\textbf{Summary for RQ2 (Impact of expertise)}: Using CAs introduced the One-pass, No-Pause-Write strategy, which was absent in the baseline. Participants accepted significantly more and ignored fewer suggestions when using the experienced CA than the novice CA. Additionally, messages sent to the experienced CA were more evenly distributed across different categories, while those to the novice CA were primarily focused on problem identification and generation.
    }
  }
}

\subsubsection{Video Analysis Strategies when Working with the CAs}

\begin{figure*}[h]
  \centering
  \resizebox{\textwidth}{!}{%
        \scalebox{1}{\includegraphics{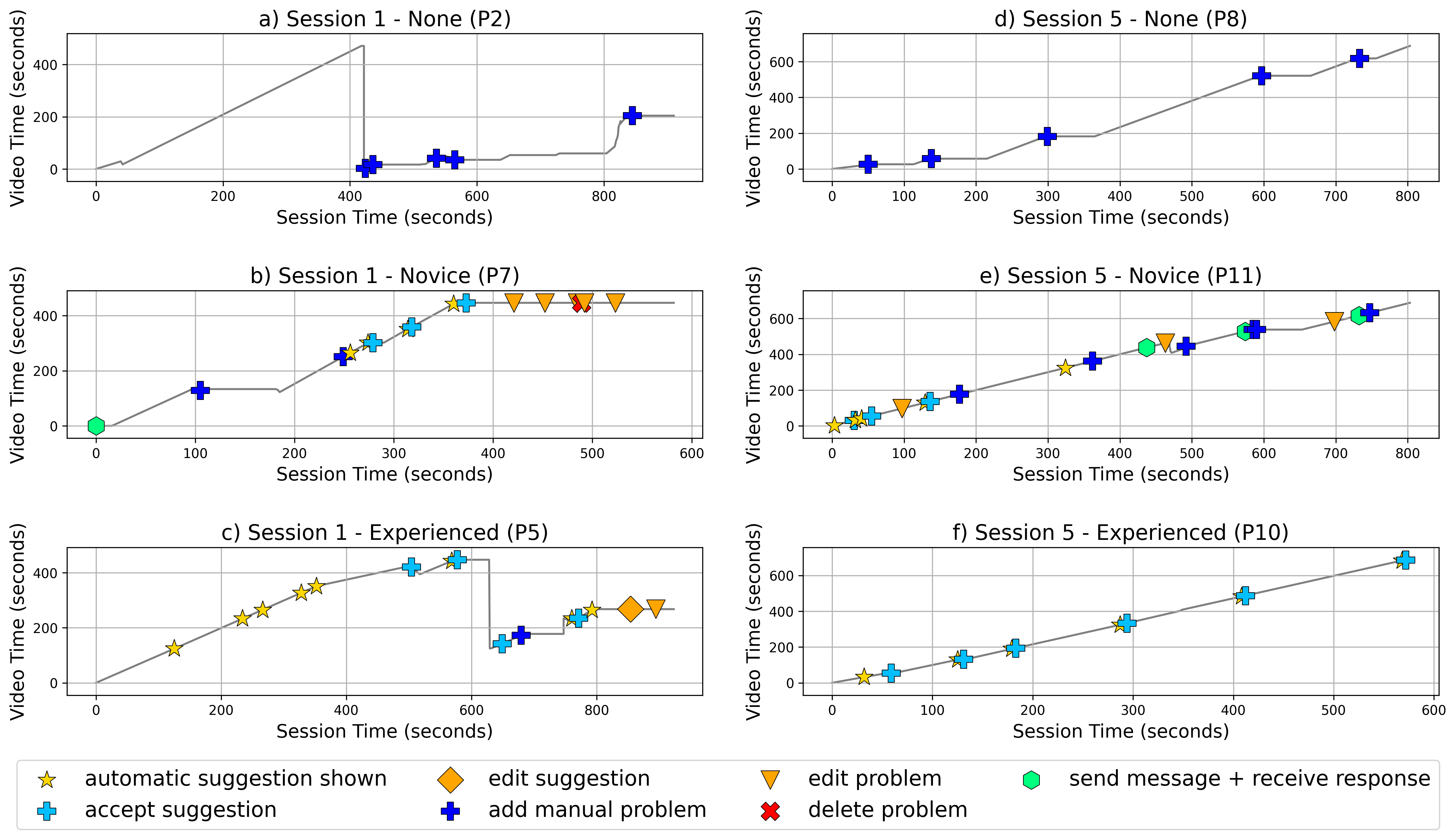}}
    }
  \caption{Timeline plots of analysis behaviors with session time on the x-axis and video time on the y-axis. The first \rv{column} shows the three conditions (no CA, novice CA, and experienced CA) during session 1, while the second \rv{column} presents these same conditions during session 5. \rv{These video timelines were chosen to illustrate the five main analysis strategies visually.}}
  \Description{Graphical representation of six session timelines, comparing video time and session time. Plot a) Session 1 - No (P2) shows a near-linear increase in video time in the first pass with pauses in the second pass. Plot b) Session 1 - novice (P7) shows consistent video time progress in the first pass, with a long pause to edit problems at the end. Plot c) Session 1 - Experienced (P5) shows frequent interactions in the first pass and second pass. Plot d) Session 5 - No (P8) shows a linear progression with less frequent interactions. Plot e) Session 5 - novice (P11) shows frequent interactions, maintaining a steady increase in video time. Plot f) Session 5 - Experienced (P10) shows a smooth, linear increase in video time with frequent user interactions.}
  \label{fig-chp7:timeline-plots}
\end{figure*}

\begin{figure*}[h]
  \centering
  \resizebox{0.8\textwidth}{!}{%
        \scalebox{1}{\includegraphics{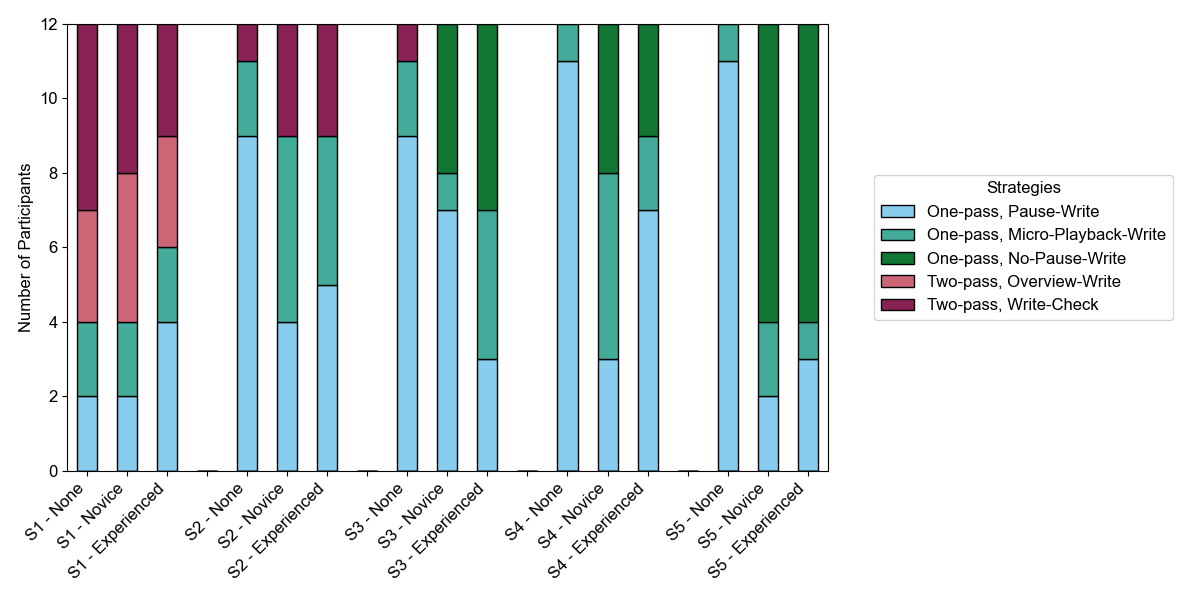}}
    }
  \caption{Stacked bar chart showing the distribution of five video analysis strategies by session and condition. Two-pass strategies were more frequent in the first two sessions, while the One-pass, No-Pause-Write strategy emerged in session 3 and was used by two-thirds of participants by session 5.}
  \Description{Bar graph titled "Video Analysis Strategies Separated By Session and CA Condition." The x-axis contains 15 bars, each representing the none, novice, and experienced conditions in sessions 1 to 5. The y-axis represents the number of participants, ranging from 0 to 12 in increments of 2. The legend shows different colors for the five strategies: One-pass, Pause-Write; One-pass, Micro-Playback-Write; One-pass, No-Pause-Write; Two-pass, Overview-Write; and Two-pass, Write-Check. The bars show varying numbers of participants across different sessions with the two-pass strategies being more frequent in the first two sessions and One-pass, No-Pause-Write emerging in session 3.}
  \label{fig-chp7:analysis-strategies}
\end{figure*}

In our study, \rv{twelve} participants reviewed \rv{15 videos, resulting in 180 recorded sessions.}
We examined how many passes they made on a video and their playback behaviors when going through a pass. 
Fig. \ref{fig-chp7:timeline-plots} shows six timeline plots of typical behaviors when analyzing one video.
\rv{These examples were selected to represent the main analysis strategies observed, rather than to show an idealized progression.}
Fig. \ref{fig-chp7:analysis-strategies} displays the distribution of the five main analysis strategies across 180 sessions: 


\textbf{1) Two-pass, Overview-Write}: This strategy involved first watching the video from beginning to end to gain an overview before going back to add problems in the second pass (Fig. \ref{fig-chp7:timeline-plots}a and Fig. \ref{fig-chp7:timeline-plots}c). 
It was employed for 10 videos (5.6\%) by four participants and only appeared in the first session.

\textbf{2) Two-pass, Write-Check}: This strategy is similar to the previous one, but participants recorded problems or responded to suggestions during the first pass and then went back to check (Fig. \ref{fig-chp7:timeline-plots}b). 
It was employed for 20 videos by seven participants in the first two sessions, and with one video in session 3 (11.1\% of all videos).

\textbf{3) One-pass, Pause-Write}: This was the most frequently employed strategy and was used by all participants, accounting for 82 videos (45.6\%).
Participants paused to record usability problems or respond to suggestions while reviewing the video once (Fig. \ref{fig-chp7:timeline-plots}d).

\textbf{4) One-pass, Micro-Playback-Write}: This strategy involved rewinding a small portion of the video before recording a problem and was employed for 36 videos (20\%) by eight participants. 
For example, Fig. \ref{fig-chp7:timeline-plots}e shows P11 rewinding between 400 to 500 seconds to review and add a problem. 
Other instances of micro-playback occurred in response to an automatic suggestion, where participants rewound the video to check whether they agreed with the suggestion. 
Fig. \ref{fig-chp7:analysis-strategies} shows that this strategy was more frequently used in the first two sessions (17 videos) than the last two (12 videos).

\textbf{5) One-pass, No-Pause-Write}: This strategy involved minimal pausing, with participants directly accepting or editing suggestions as they appeared. 
Fig. \ref{fig-chp7:timeline-plots}f shows a straight line where P10 accepted all suggestions. 
This strategy was applied to 32 videos (17.8\%) by eleven participants, but it only appeared in conditions with CAs, as the baseline condition required pausing to type problem descriptions. 
It emerged in the last three sessions, accounting for two-thirds of videos by session 5.

\subsubsection{Responses to Automatic Suggestions from the CAs}

In this section, we describe the three types of responses to automatic suggestions: accept, edit, or ignore.

\begin{figure}[h]
  \centering
  \begin{subfigure}[h]{0.48\textwidth}
    \centering
    \includegraphics[width=\textwidth]{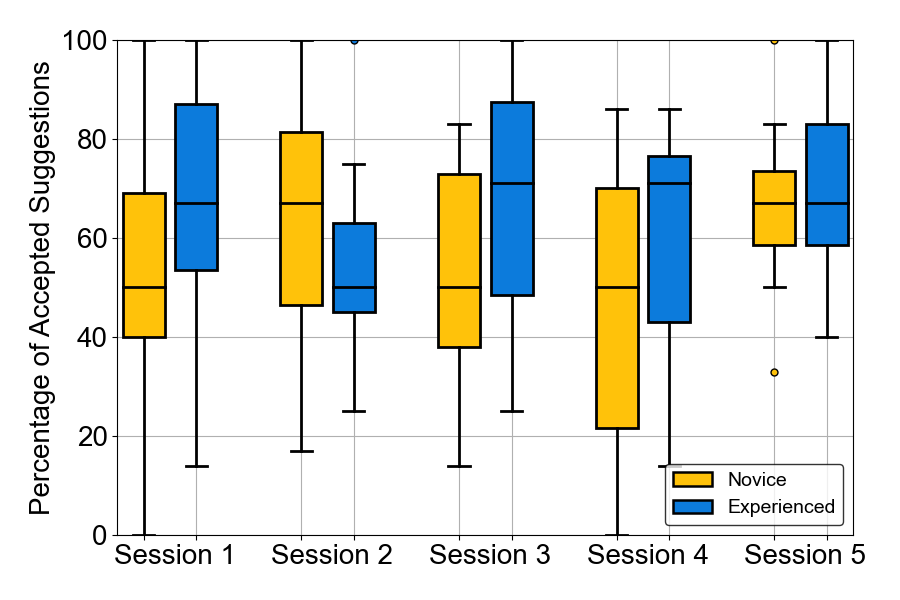}
    \caption{Acceptance rates of novice and experienced CA's suggestions.}
    \label{fig-chp7:accept-rates}
  \end{subfigure}
  \hfill
  \begin{subfigure}[h]{0.48\textwidth}
    \centering
    \includegraphics[width=\textwidth]{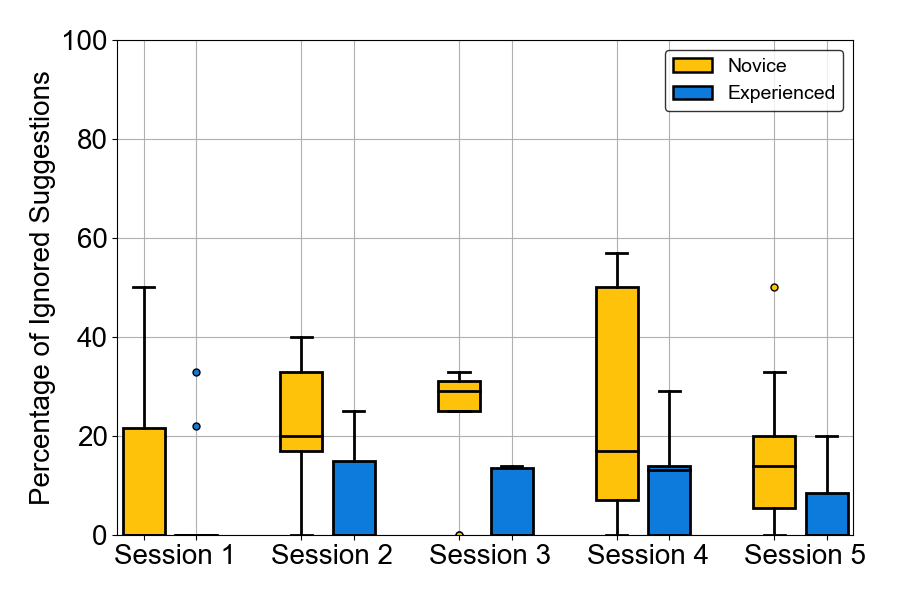}
    \caption{Ignore rates of novice and experienced CA's suggestions.}
    \label{fig-chp7:ignore-rates}
  \end{subfigure}
  \caption{Box plots displaying the percentage of (a) accepted suggestions and (b) ignored suggestions, separated by session and CA condition. Acceptance rates were generally higher for the experienced CA compared to the novice CA, while ignore rates were higher for the novice CA. Additionally, ignore rates for both CAs increased after session 1.}
  \Description{Two box plots side by side. Each plot contains a y-axis representing the percentage of suggestions, ranging from 0 to 100, and an x-axis with five pairs labeled from Session 1 to Session 5. Each session has two conditions: novice and Experienced. Plot a) shows the acceptance rates with the boxes for 'Experienced' resting higher than those for 'Novice' while Plot b) shows the ignore rates with the boxes for 'Experienced' being lower than those for 'Novice'.}
\end{figure}

\paragraph{Rate of Accepted Suggestions}
We found that only the effect of \textit{condition} was significant\rv{, meaning that }
using the CA over time did not significantly impact how much participants accepted the suggestions. 
For the comparison between CA conditions, participants accepted significantly more suggestions from the experienced CA ($M = 66.5\%, SD = 23.6\%$) than the novice CA ($M = 56.0\%, SD = 26.0\%$) throughout all sessions. 
Fig. \ref{fig-chp7:accept-rates} shows a box plot of the percentage of accepted suggestions separated by session and CA condition, where the boxes for experienced CA generally rested higher than those of the novice CA. 
Participants noted that the experienced CA provided more detailed and specific suggestions, while the novice CA offered general observations while providing superficial causes (e.g., stating that the user expressed frustration without identifying its underlying cause). 
For instance, although both CAs identified the same problem in the VR game, the novice CA's suggestion to ``add a tutorial'' was seen as too vague and failed to recognize that a tutorial already existed but was difficult for the user to find (P2). 
In contrast, the experienced CA offered specific recommendations, such as making the tutorial button more noticeable by increasing its size and contrast against the background and suggesting exact text edits. 
As a result, the experienced CA's suggestions were accepted more often than those of the novice CA.

\paragraph{Rate of Edited Suggestions}
There were no significant patterns for the edit rate, suggesting that how much participants edited the suggestions did not change over time or for different conditions (shown in Fig. \ref{fig-chp7:edit-rates} in the Appendix).

\paragraph{Rate of Ignored Suggestions}
Both the effect of \textit{session} and \textit{condition} were significant for the ignore rate. 
Post-hoc pairwise comparisons between session pairs showed that the ignore rates in session 2 ($M = 16.2\%, SD = 13.6\%$), session 3 ($M = 13.3\%, SD = 13.8\%$), and session 4 ($M = 17.8\%, SD = 19.0\%$) were significantly higher (all $p < .05$) than the ignore rates in session 1 ($M = 8.0\%, SD = 15.4\%$).
The difference in ignore rates demonstrates a novelty effect where participants were less likely to ignore suggestions at the beginning. 
From session 2 onward, participants reported that they began to recognize the limitations of the CAs, and as a result, started ignoring more suggestions.
Fig. \ref{fig-chp7:ignore-rates} shows a box plot of the percentage of ignored suggestions separated by session and CA condition. 
The boxes for experienced CA are always lower than the novice CA since participants ignored 19.7\% of all suggestions from the novice CA while only 6.3\% of suggestions from the experienced CA. 
Many participants felt that the novice CA often provided irrelevant suggestions (false positives), which aligns with our evaluation in Section \ref{sec:performance-evaluation}. 
P3 noted that it felt too sensitive, often flagging minor delays as problems.
For instance, P8 believed that a user's difficulty completing a task was due to unfamiliarity with the app, not a problem with the app itself. 
Similarly, P5 interpreted a user's statement of ``I do not like this'' as a personal preference rather than an app problem.
As a result, participants disregarded these suggestions, leading to higher ignore rates for the novice CA.

\subsubsection{Categories of Messages Sent to the CAs}

In total, participants sent 170 messages to the CAs, and we identified 17 different categories of messages.
Table \ref{tab:message-cateogories} in the Appendix displays the categories, frequency, and example messages. 

\begin{figure*}[h]
  \centering
  \begin{subfigure}[h]{0.38\textwidth}
    \centering
    \includegraphics[width=\textwidth]{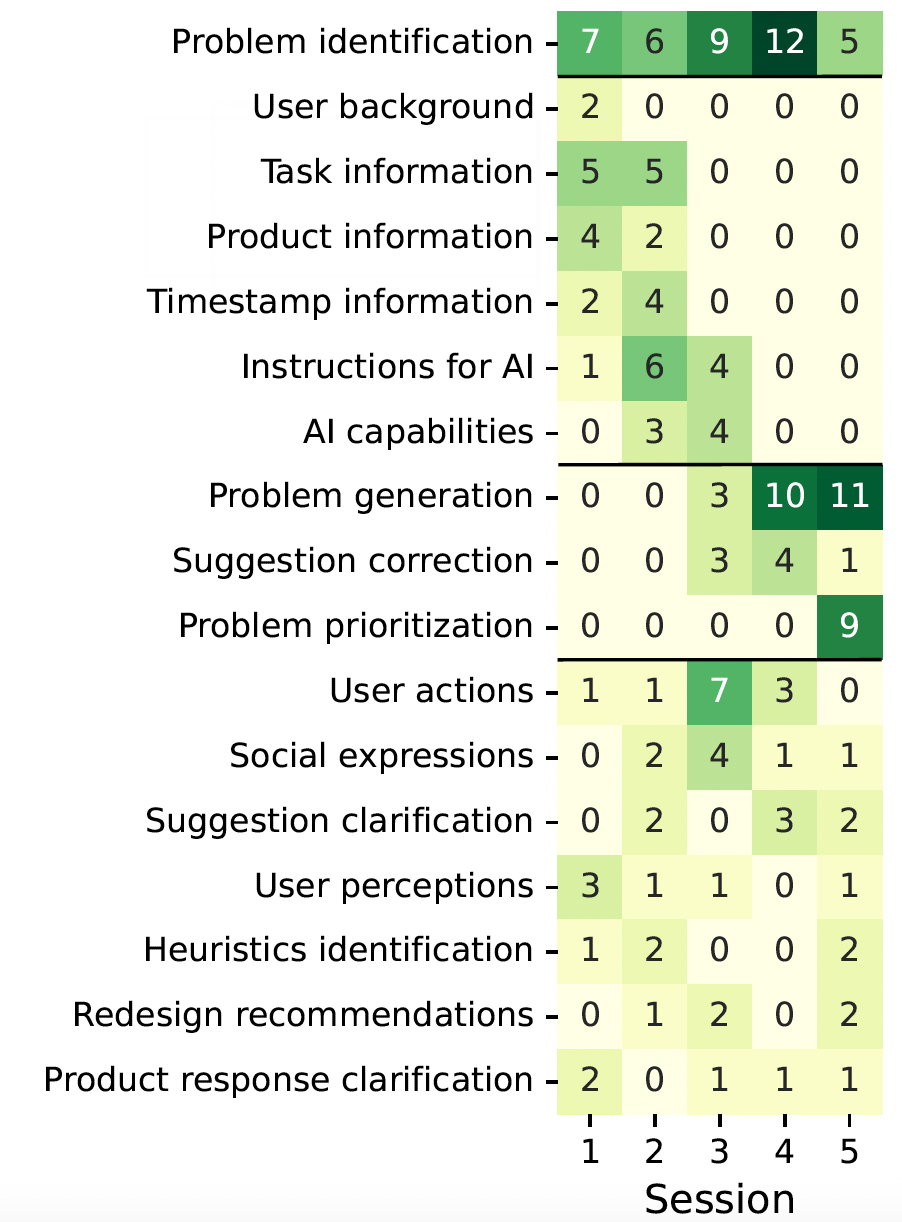}
    \caption{Message categories by session.}
    \label{fig-chp7:messages-heatmap}
  \end{subfigure}
  \hfill
  \begin{subfigure}[h]{0.48\textwidth}
    \centering
    \includegraphics[width=\textwidth]{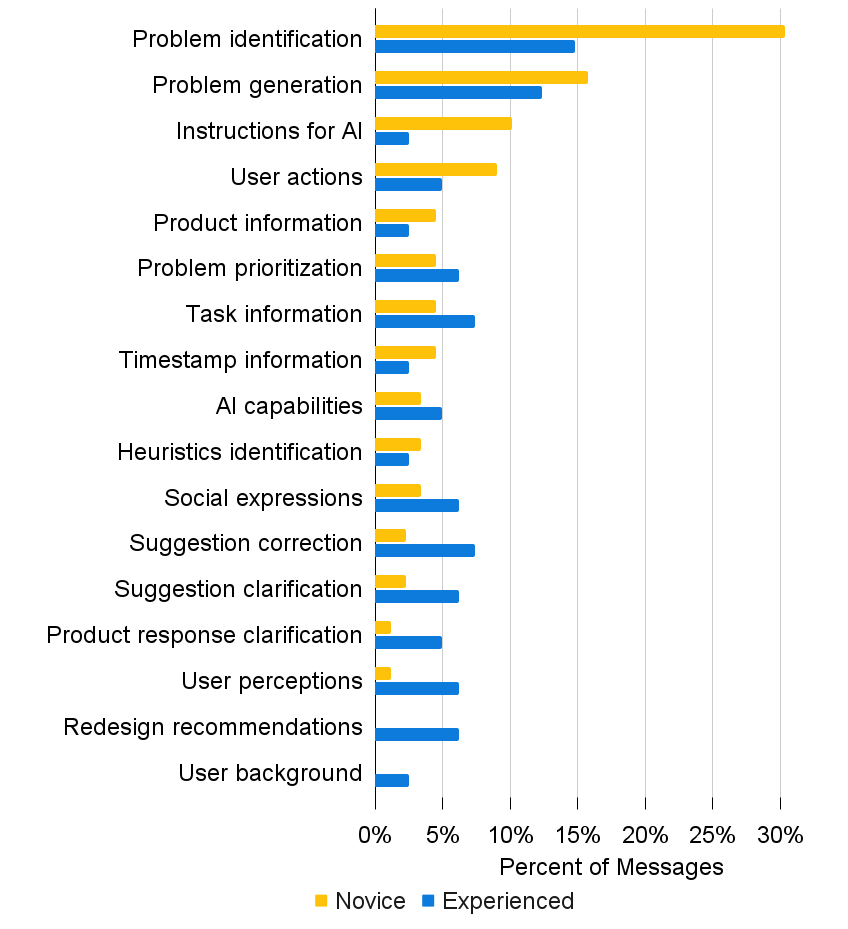}
    \caption{Message categories by CA.}
    \label{fig-chp7:messages-by-ca}
  \end{subfigure}
  \caption{Visualizations of message categories: (a) Heatmap by session: Problem identification appeared in all sessions. The next six categories were present only in the first three sessions, and the next three categories only in the last three sessions. The remaining categories showed no clear pattern. (b) Bar chart by CA: Message categories were more evenly distributed for the experienced CA.}
  \Description{On the left: heatmap containing 6 columns and 19 rows: the first column lists the 18 categories of messages in the order of most frequent to least, and the next five columns show the number of messages in that category for each of the five sessions. On the right: Bar chart containing 36 rows, with the yellow bars representing novice CA and the blue bars representing experienced CA. The y-axis shows categories of messages, and the x-axis indicates the frequency of each category. The row lengths for the novice CA range from 0 to 30, while the row lengths for the experienced CA range from 2 to 15.}
\end{figure*}

\paragraph{Impact of Longitudinal Use}
Since there were no significant differences in the number of messages between sessions, we focused on analyzing trends in message categories and their distribution across sessions (Fig. \ref{fig-chp7:messages-heatmap}).
The category with the highest number of messages (39 in total) was 
\textbf{problem identification}. 
In this category, participants questioned whether the CA had missed any usability problems or if additional problems could be identified, indicating that this was a primary concern during the analysis. 
This was the only message category that appeared in all five sessions.

In the initial sessions, participants tended to ask simpler questions, such as requesting information on tasks, products, and users. 
They also explored the CA's capabilities by inquiring whether it could generate a problem, identify causes, or provide redesign recommendations based on the information they supplied. 
In the later sessions, participants increasingly relied on the CA for more complex tasks, such as generating precise wording for identified problems and prioritizing the list of usability problems after analyzing the video. 
They also engaged in more dialogue turns, such as informing the CA of additional causes it had missed and seeking clarification on specific recommendations.


\paragraph{Impact of Varying UX Experience}

Fig. \ref{fig-chp7:messages-by-ca} shows the categories and number of messages sent to the novice and experienced CAs. 
Participants sent 89 messages to the novice CA and 81 messages to the experienced CA, with no significant difference between the two conditions.
\rv{However, the content of these messages were different.}
For the novice CA, the messages mainly concentrated on two categories: identifying usability problems (30.3\%) and generating problem descriptions (15.7\%), \rv{reflecting more task-oriented exchanges}. 
\rv{In contrast,} message to the experienced CA were more evenly distributed and included two categories that did not occur for the novice CA: redesign recommendations and user background, \rv{indicating more strategic and reflective interactions}.
\rv{Participants also engaged more in user perception and suggestion clarification and correction messages with the experienced CA, suggesting greater willingness to discuss higher-level design rationale and to negotiate the AI's output. 
This pattern implies that as perceived AI expertise increased, participants shifted from basic information extraction toward collaborative reasoning.}


\rv{
\subsubsection{Individual Differences in Analysis Behaviors Among Participants}

Since our sample was small, we did not compute correlations between participants' UX backgrounds or AI usage habits and their behaviors. However, we observed several qualitative trends.
Participants who used the ``Two-pass, Overview-Write'' strategy tended to be more meticulous, had less UX experience (2-4 years), and were only somewhat familiar with AI. For example, P2, P5, and P7 carefully validated the CAs' suggestions and reviewed video context before writing, using this strategy across all three videos in their first session. As they became more familiar with the CAs, they shifted to faster approaches.
In contrast, P10 and P12, who had more UX experience (7-8 years), higher familiarity with usability analysis (rated 4-5), and frequent AI use (rated 5), immediately adopted the ``One-pass, Pause-Write'' strategy and even sped up videos in the first session.

We also observed differences in message categories. 
Problem prioritization emerged only in the final session and was primarily used by P12, who described prioritization as part of her normal workflow when completing the analysis. 
Problem generation was common among P5, P6, and P11 once they discovered that CAs could help articulate findings. For instance, P11 prompted the novice CA with: \textit{``Provide the problem, cause, and redesign recommendation for the user's confusion about reload.''} In interviews, P11 explained that the CA sometimes missed issues but was faster at writing out the descriptions once she identified the problem.

Acceptance rates also varied significantly across participants for both the novice CA ($F_{11, 48} = 2.62, p < .05, \eta_{p}^{2} = 0.38$) and expert CA ($F_{11, 48} = 2.81, p < .01, \eta_{p}^{2} = 0.39$), based on one-way between-subjects ANOVA. 
Participants with the highest acceptance rates were P1-P3, who each had 3 years of UX experience and reported being only slightly familiar with AI. In contrast, participants with the lowest acceptance rates (P9-P11) reported higher AI familiarity; for example, P10 rated himself as ``5 - extremely familiar.'' Prior work suggests that experts may be more hesitant to adopt AI because deeper technical understanding often heightens awareness of system limitations \cite{simkute_it_2024}.
}

\subsection{Analytic Performance}
\label{sec:analytic-performance}

\fbox{
  \parbox{0.97\linewidth}{
    \textit{\textbf{Summary for RQ1 (Change over time)}: Participants identified more problems in the final session when using the novice and experienced CAs, indicating that familiarity with the videos and CAs improved their analysis.}
    
    \textit{\textbf{Summary for RQ2 (Impact of expertise)}: Using CAs significantly increased the total number of problems, unique problems, problem coverage, and inter-rater reliability (Fleiss' kappa) compared to the baseline. The experienced CA produced significantly better results than the novice CA.}
  }
}


\subsubsection{Number of Usability Problems}

\begin{figure}[h]
  \centering
  \begin{subfigure}[h]{0.48\textwidth}
    \centering
    \includegraphics[width=\textwidth]{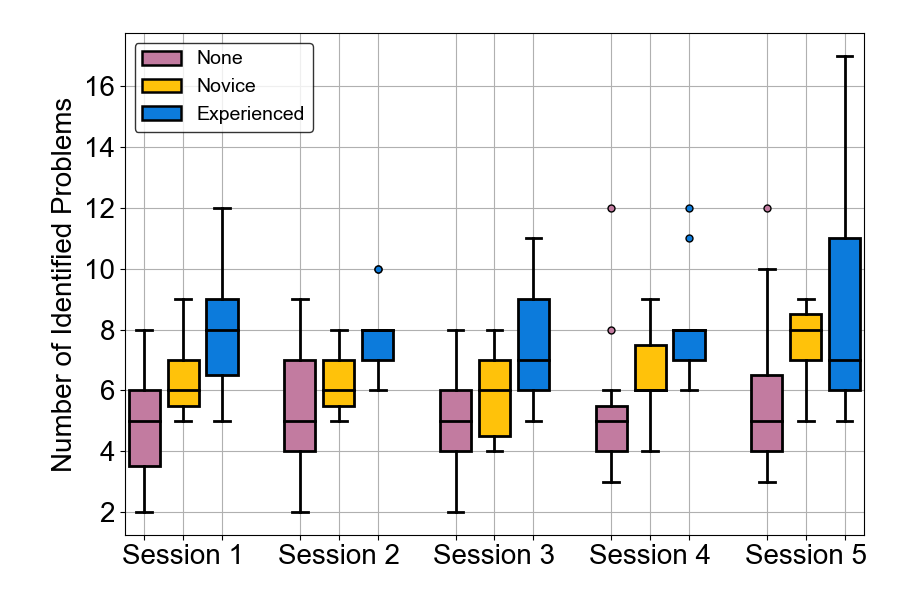}
    \caption{Number of identified problems.}
    \label{fig-chp7:total-problems}
  \end{subfigure}
  \hfill
  \begin{subfigure}[h]{0.48\textwidth}
    \centering
    \includegraphics[width=\textwidth]{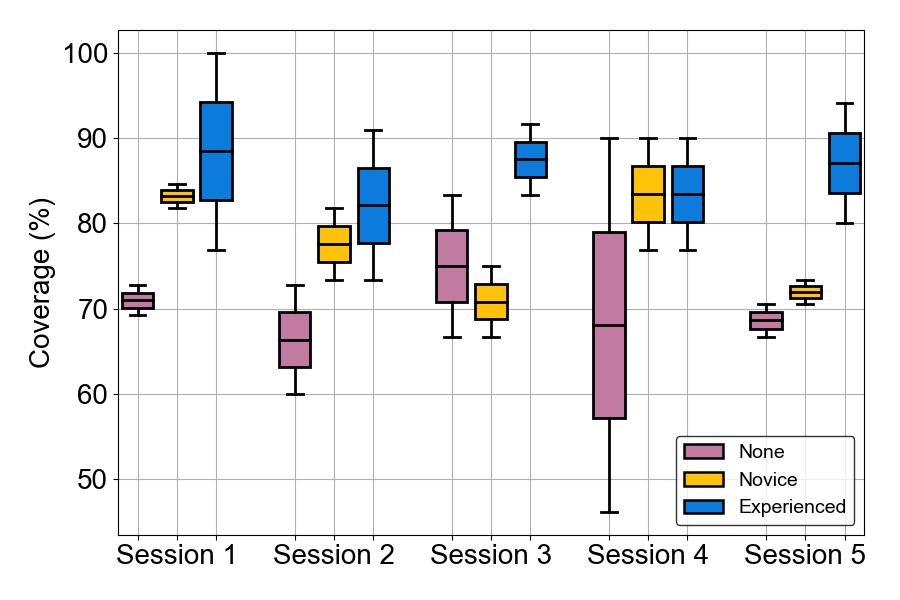}
    \caption{Coverage of unique problems.}
    \label{fig-chp7:coverage}
  \end{subfigure}
  \caption{Box plots show (a) Total number of identified problems: Problems were highest in session 5, and were higher with the experienced CA, followed by the novice CA, and least with no CA. (b) Coverage of unique problems: There were no session-based patterns, but the experienced CA achieved higher coverage compared to both novice CA and no CA.}
  \Description{Two box plots side by side. Each plot contains an x-axis with five pairs labeled from Session 1 to Session 5. Each session has three conditions: No, Novice, and Experienced. Plot a) has a y-axis representing the number of problems, ranging from 0 to 16 and the boxes for the `Experienced' condition lie higher than those for 'Novice', which rest higher than those for 'No'. Plot b) has a y-axis representing the percent coverage of problems, ranging from 0 to 100 and the boxes follow a similar trend as Plot a).}
\end{figure}

In total, participants identified 1217 problems from the 15 usability videos: 330 on their own, 393 with the novice CA, and 494 with the experienced CA. 
Fig. \ref{fig-chp7:total-problems} shows the total number of identified problems per video separated by session and condition, while Table \ref{tab-chp7:stat-tests} shows the statistical test results. 
Both the effect of \textit{session} and \textit{condition} were significant.
Post-hoc pairwise comparisons between session pairs showed that the number of identified problems in session 5 ($M = 7.5, SD = 3.1$) was significantly higher ($p < .01$) than in session 3 ($M = 6.3, SD = 2.0$).
This difference could be attributed to participants being already familiar with the common usability problems encountered by session 5 and relying more on the CAs at the end. 

\begin{table}[h]
  \centering
  \caption{Analytic performance measures based on condition, reported in ``Mean (Standard Deviation)''}
  \label{tab-chp7:analytic-performance}
  \small
  \begin{tabular}{p{2.4cm} p{1.5cm} p{1.4cm} p{2cm}}
    \toprule
    \textbf{Variable} & \textbf{No CA} & \textbf{Novice CA} & \textbf{Experienced CA} \\
    \midrule
    Total problems per participant per video & 
    5.5 (2.4) & 
    6.6 (1.4) & 
    8.2 (2.5) \\
    \midrule
    Unique problems per video & 
    8.6 (1.6) & 
    9.3 (1.6) & 
    10.3 (2.3) \\
    \midrule
    Problem coverage per video & 
    70.9\% (13.3\%) & 
    76.3\% (7.8\%) & 
    84.3\% (8.3\%) \\
    \midrule 
    Fleiss' kappa per video & 
    0.40 (0.17) & 
    0.58 (0.18) & 
    0.62 (0.15) \\
    \bottomrule
\end{tabular}
\end{table}

Table \ref{tab-chp7:analytic-performance} shows the mean and standard deviation of the number of problems identified per participant based on condition.  
Post-hoc pairwise comparisons showed that the number of problems found by the participants using the novice and experienced CAs was significantly higher than no CA assistance ($p < .01$ and $p < .0001$, respectively). 
Furthermore, there were significant differences between CA conditions, where participants using the experienced CA identified significantly more problems than the novice CA ($p < .0001$).
This shows that using the experienced CA led to the best results in the number of identified problems. 


\subsubsection{Number of Unique Problems and Coverage}

In addition to counting the total number of problems, we also examined the number of unique problems for each video (Table \ref{tab-chp7:analytic-performance}). 
\rv{As mentioned in Section \ref{sec:procedure}, four participants reviewed each video-condition combination. 
The number of unique usability problems was then calculated as the total number of distinct problems identified across those four participants.}
Only the effect of \textit{condition} was significant. 
Post-hoc pairwise comparisons showed that the number of unique problems with the experienced CA was significantly higher than without any CA assistance, with $p < .05$. 
The novice CA was not significantly different than the other two conditions. 

We also calculated the coverage of each condition by dividing the number of unique problems by the total unique problems across all twelve participants for each video (Fig. \ref{fig-chp7:coverage}).
Similar to unique problems, only the effect of \textit{condition} was significant. 
Post-hoc pairwise comparisons showed that the coverage with the experienced CA was significantly higher than without any CA assistance ($p < .01$) and the novice CA ($p < .05$).

\subsubsection{Inter-rater Reliability}

\begin{figure}[h]
    \centering
    \includegraphics[width=0.48\textwidth]{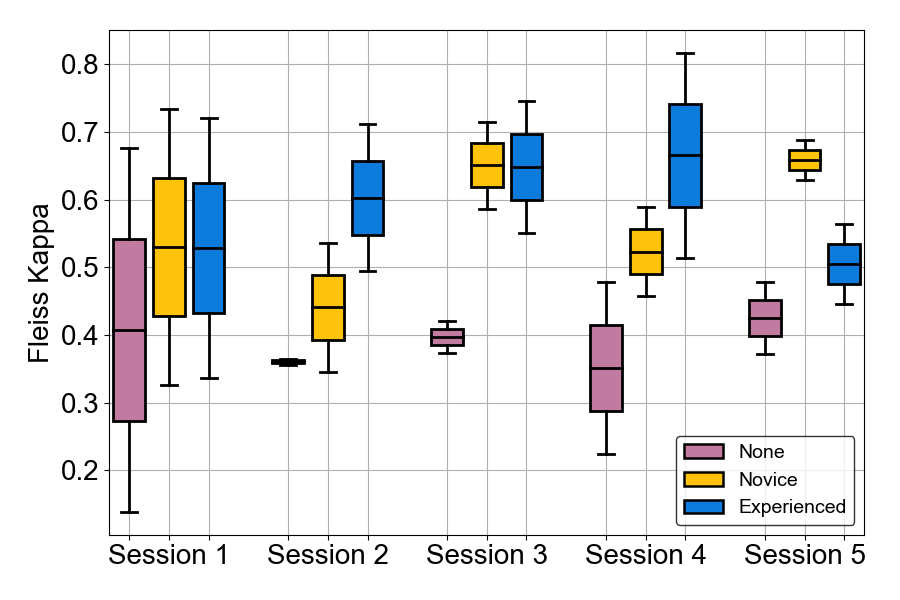}
    \caption{Box plot showing Fleiss' kappa separated by session and CA condition.}
    \label{fig-chp7:IRR}
    \Description{A box plot titled ``Fleiss' Kappa Separated by Session and CA Condition.'' The y-axis represents kappa, ranging from 0 to 0.8. The x-axis lists fifteen labels, showing five sessions, each with three conditions: No, Low, and High. The boxes show the median, quartiles, and range of kappa with dots representing outliers. The boxes for 'No' rest lower than those for 'Low' and 'High'.}
\end{figure}

In line with prior work using Fleiss' kappa as a measure of inter-reliability between ratings from ChatGPT and humans \cite{khademi_can_2023}, we also computed this value based on the results from the four participants in each condition for each video.
Fig. \ref{fig-chp7:IRR} shows a box plot of Fleiss' kappa separated by session and condition. 
Only the effect of \textit{condition} was statistically significant. 
Post-hoc pairwise comparisons showed that values in the novice and experienced CA were significantly higher than without any CA assistance ($p < .01$ and $p < .001$ respectively).
Interpreting the kappa values from Landis and Koch's table \cite{landis_measurement_1977}, our findings indicate there was moderate agreement (0.4 - 0.6) for the no CA and novice CA conditions, while experienced CA showed substantial agreement (>0.6).

\subsection{Subjective Feedback on Human-AI Collaborative Analysis}
\label{sec:subjective-feedback}

\fbox{
  \parbox{0.97\linewidth}{
    \textit{\textbf{Summary for RQ1 (Change over time)}: For both CAs, the novelty effect caused a drop in perceived efficiency and trust in session 2 compared to session 1, which recovered from session 3 onward. Participants began to notice differences in UX expertise between the two CAs from session 2 onward.}
    
    \textit{\textbf{Summary for RQ2 (Impact of expertise)}: Participants rated the experienced CA as significantly more efficient, trustworthy, and complete in its suggestions, which increased their confidence. Overall, the experienced CA was preferred.}
  }
}

\begin{figure*}[h]
  \centering
  \resizebox{\textwidth}{!}{%
        \scalebox{1}{\includegraphics{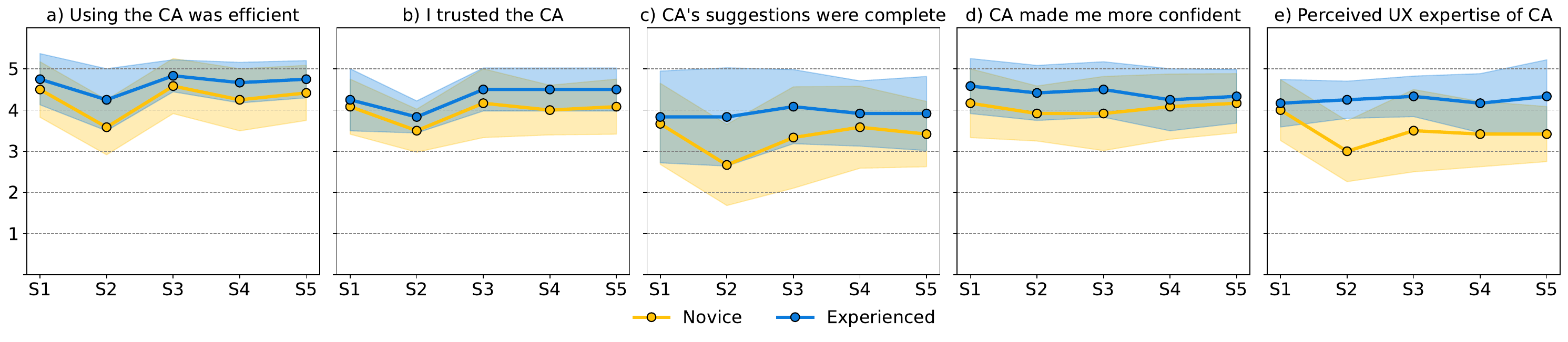}}
    }
  \caption{Line charts displaying the mean (circular markers) and standard deviation (shaded area) of Likert ratings for a) perceived efficiency, b) trust, c) suggestion completeness, d) analysis confidence, and e) UX expertise over five sessions. Statements a-d were rated from 1 (strongly disagree) to 5 (strongly agree), while e was rated from 1 (none) to 5 (expert). The novice CA saw drops in efficiency, trust, and completeness in session 2, while the experienced CA showed declines in efficiency and trust. The perceived UX expertise of the novice CA declined starting in session 2.}
  \Description{Chart 1 shows "Using the CA was efficient"; Chart 2 shows "I trusted the CA"; Chart 3 shows "Suggestions were complete"; Chart 4 shows "CA made me more confident"; Chart 5 shows "Perceived UX expertise of CA". The novice CA saw drops in efficiency, trust, and completeness in session 2, while the experienced CA showed declines in efficiency and trust. The perceived UX expertise of the novice CA declined starting in session 2.}
  \label{fig-chp7:ratings-perceptions}
\end{figure*}

Fig. \ref{fig-chp7:ratings-perceptions} shows the ratings on a 5-point Likert scale for the efficiency of using the CA, trust in the CA, perceived completeness of suggestions, confidence in analysis, and perceived UX expertise of the CA.


\paragraph{Efficiency of Collaborating with the CA}
Overall, participants felt that using the CAs was very efficient for their analysis ($Md = 5, IQR = 1$). 
P12 mentioned, \textit{``it was very efficient and made me feel like I was getting through analysis much quicker without typing everything myself.''}
When comparing participants' ratings across sessions and conditions, both factors led to significant differences. 
Post-hoc pairwise comparisons between session pairs showed that the efficiency ratings in session 2 were significantly lower than the other four sessions.
In session 2, participants began to notice the CA's limitations, indicated by asking more questions about its capabilities and starting to figure out how to best collaborate with the CA. 
For instance, P2 mentioned in session 2's interview: \textit{``I noticed a problem and waited for the AI suggestion so I wouldn't have to type it, but when it didn't detect the problem, I had to type it manually and even rewatch the segment to frame the statements accurately.''}
Overall, participants rated the experienced CA as significantly more efficient than the novice CA 
since the novice CA often missed problems, requiring participants to add them manually. 
P9 noted, \textit{``It didn't catch the things I was catching. Sometimes, multiple usability issues occurred in succession, but it only identified the first one, so I had to type the rest in myself, which took more time.''}
In contrast, in the last 2-3 sessions, participants using the experienced CA increased their efficiency by speeding up the videos. P1 explained, \textit{``I actually sped up the video because I know it will point out the right things. I don't double check the suggestions from the [experienced CA], but I always check for the [novice CA]''}-P3.

\paragraph{Trust in the CA}
Overall, participants somewhat agreed that they trusted the information provided by both CAs ($Md = 4, IQR = 1$). 
The trust ratings were significantly different depending on the \textit{session} and ``condition.''
Similar to the efficiency ratings, post-hoc pairwise comparisons between session pairs showed that the trust ratings in session 2 were significantly lower than the other four sessions (\textit{all Md = 4, IQR = 1}).
This difference shows an overall trend where both the perceived efficiency and trust were high in session 1, decreased in session 2, and then recovered from session 3 onwards. 
P4 explained the reasons behind this trend: \textit{``I was very impressed in the first session, but I got more critical once I noticed mistakes in session 2. After session 3, I knew exactly what to expect from the videos and the assistant, so my trust and speed increased until the end.''}
Conversely, other participants felt skeptical during the first session, then their trust gradually increased throughout the sessions because \textit{``the more correct examples I saw, the more I trusted it''}-P6. 

The levels of trust were also significantly different depending on the condition. 
P10 mentioned that their \textit{``trust in [experienced CA] increased, but decreased for the [novice CA].''}
Similarly, P12 noticed that novice CA did not always provide appropriate redesign recommendations, so they had to fill in these missing details, decreasing their overall reliance on this version. 

\paragraph{Perceived Completeness of Suggestions}
Participants agreed that the CAs provided a complete list of suggestions overall ($Md = 4, IQR =1$). 
Only the effect of \textit{condition} led to significantly different ratings, indicating that participants' opinions on completeness did not change over time. 
However, they noticed that the experienced CA provided significantly more complete suggestions than the novice CA. 
In particular, P7 found the novice CA's suggestions helpful when they appeared but felt they were often missing. She emphasized the importance of consistency, which lowered her trust in the novice CA, stating, \textit{``I had to be more active and alert because I knew it wasn't reliable.''}

\paragraph{Confidence in Analysis}
Participants generally agreed that the CAs increased their confidence in usability analysis ($Md = 4, IQR = 1$). 
P11 described the CAs as \textit{``a second opinion, which is always nice to have. I'm not afraid to disagree; I have the power to decide whether to accept it, but it could be especially helpful when I'm unsure.''}
Although participants' ratings for confidence remained consistent over time, they reported feeling significantly more confident when using the experienced CA compared to the novice CA.

\paragraph{Perceived UX Expertise}
Participants rated the CA's level of UX expertise from 1 (none) to 5 (expert). 
The ratings were significantly different between sessions and conditions. 
Interestingly, the interaction effect was also significant, indicating that the two variables of session and condition depend on each other. 
For the novice CA, post-hoc pairwise comparisons between session pairs showed that ratings in session 2 were significantly lower than in session 1.
For the experienced CA, the median ratings were consistent between 4 and 5, so there were no significant differences over time. 
Fig. \ref{fig-chp7:ratings-perceptions}e shows that the ratings for both CAs were nearly the same in session 1, but starting from session 2, the ratings for the novice CA became lower than those of the experienced CA, suggesting that participants started noticing the differences after the first session. 
In the final interview, participants described the novice CA as an \textit{``intern''} or \textit{``entry-level''} UX evaluator, while they felt that the experienced CA was \textit{``entry-level,''} \textit{``mid-level,''} or \textit{``expert.''} 

\paragraph{Overall Comparison of Novice and Experienced CAs}
In the final interview, participants assigned an overall satisfaction rating from 1 to 10 to each CA. 
The novice CA received $Md = 7.5, IQR = 1.25$, while the experienced CA achieved $Md = 9, IQR = 0.25$. 
A one-way ANOVA showed that the difference was significant, demonstrating a clear preference for the experienced CA. 
P8 mentioned that the experienced CA \textit{``made me more efficient and caught more of the problem than the other version''}. 
While most participants echoed this sentiment, some still found the novice CA valuable for specific use cases. 
P10 mentioned that \textit{``it would be good for training new researchers and interns since they can learn effectively with someone at their own level. This way they can apply their skills and judge how well the AI is doing without overdepending on the AI since it's not that good.''}
Overall, our study revealed key differences in analytic behaviors and perceptions between novice and experienced CAs, which we elaborate on in the following section.

\section{Discussion}

This section expands on the impact of multi-session analysis, discussing the implications for longitudinal studies in human-AI collaboration. 
Additionally, it discusses practical use cases for both novice and experienced CAs.

\subsection{Impact of Multi-Session Analysis (RQ1)}

\subsubsection{Evolution of Video Analysis Behaviors}

We identified five main video analysis strategies, consistent with Fan et al.'s findings on using visualizations to flag usability problems \cite{fan_vista_2020}. 
\rv{In Fan et al.'s study, participants analyzed three videos in one session, with two-pass strategies common early on. 
Similarly, our participants initially used time-intensive two-pass strategies.}
By session 3, however, the one-pass, no-pause-write strategy emerged, and by session 4, two-pass approaches disappeared. 
This progression reflects participants' growing expertise: as they became more familiar with recurring usability issues and more comfortable working with the CA, their behavior shifted toward the linear, streamlined strategies typical of expert designers \cite{chen_behaviors_2022}.


\subsubsection{Changes in Attitudes Toward AI}
\label{sec:discussion-attitude-changes}


\rv{
Participants' perceived efficiency and trust in the CAs were high in Session 1 but declined in Session 2, suggesting a \textit{novelty effect}. 
Similar declines have been observed in human-robot interaction, where enthusiasm often fades as users gain more experience—for example, children's interest in a dancing robot dropped over three months \cite{tanaka_daily_2006}, families' excitement about a Roomba decreased over six months \cite{sung_robots_2009}, and the perceived reliability of a pharmacy robot declined after fifteen months \cite{hogan_factors_2020}. 
Another possible explanation is that prior evidence shows \textit{trust often declines over time} as users apply greater scrutiny to AI systems \cite{kahr_understanding_2024, dietvorst_algorithm_2015, hoffman_trust_2013}. 
This may also explain the sharper drop in perceived expertise for the novice CA, who made more early mistakes (e.g., omitting a usability problem) than the experienced CA, which is an important factor given that first impressions are crucial for trust formation \cite{siau_building_2018} and that failures exert stronger negative effects on trust than successes \cite{yu_user_2017}.

After Session 2, trust and efficiency ratings increased again, indicating a familiarization effect, in which users adapt to and internalize new technologies as they gain experience \cite{rodrigues_gamification_2022}.
Prior work shows that as users settle into stable mental models, trust judgments become less volatile over time \cite{yu_user_2017}, which may explain why the ratings became stable between sessions 3 to 5. 
Our results mirror the trajectory reported in gamification systems: an initial dip followed by recovery \cite{rodrigues_gamification_2022}, suggesting this pattern may generalize across technology domains. 
A recent longitudinal study of generative AI for science communication similarly found that familiarization unfolded over several sessions \cite{long_not_2024}, though differences in session length (20 min vs. 90 min here) limit direct comparison. 
Consistent with that work, participants in our study used the familiarization period to probe the CAs' capabilities, which aligned with the emergence of message categories such as ``AI capabilities'' and ``problem generation'' and ultimately improved collaboration.

\textbf{Caveats on Interpreting Trust Ratings.} 
Some participants recognized differences between the CA conditions by the second or third session, meaning changes in trust may reflect both adaptation and perceived system differences. 
In addition, trust was assessed primarily through self-reports, which capture deliberate attitudes but overlook implicit behaviors and remain vulnerable to social desirability bias \cite{gulati_trust_2024, hoffman_trust_2013}.
For example, users may report trusting an AI while behaviorally avoiding its suggestions \cite{gulati_trust_2024, bucinca_proxy_2020}. 
We included acceptance rates as a behavioral proxy, but the highest rates occurred in Session 5---possibly due to increased trust, accumulated familiarity with the task, task saturation, or participant fatigue---and differences were not significant. 
Future work should incorporate richer behavioral indicators (e.g., intervention frequency, response times, eye-tracking) and physiological measures (e.g., heart rate variability, skin conductance, EEG), which have been shown to approximate trust \cite{wong_trust_2025, khawaji_using_2015, ajenaghughrure_predictive_2019}. 
Despite having limitations, self-report scales, especially validated instruments such as the Trust in Automation Scale (TIAS), remain practical and widely used for measuring trust in human-AI studies \cite{mcgrath_measuring_2025}.
}

\subsubsection{Implications for Future Longitudinal Studies on Human-AI Collaboration}

We found that perceived UX expertise ratings significantly differed between the novice and experienced CAs overall, but these differences were also shaped by \textit{session}. 
In the first session, participants rated both CAs similarly, showing no clear distinction in perceived expertise. 
However, across subsequent sessions, ratings for the novice CA declined, while the experienced CA maintained consistently higher scores.
This pattern highlights that users' initial perceptions of AI systems can be overly optimistic, shaped by novelty, rather than by accurate assessments of capability. 
If our study relied on a single-session evaluation, which is common in HCI work, we might have drawn the misleading conclusion that CA expertise does not impact user perception. Instead, our longitudinal approach revealed how these perceptions evolve with continued interaction, as users recalibrate their expectations of the AI over time.
These findings underscore the necessity of longitudinal methods in human-AI collaboration research. 

\rv{
\textbf{Reflections and Lessons Learned.}
Conducting a multi-session, longitudinal study also offered several methodological lessons. 
Sustaining participant engagement over time required balancing task familiarity with novelty. 
Introducing new videos in each session, while keeping the CAs the same, helped maintain interest without altering the study design. 
Because the participants were professionals with full-time jobs, flexibility in scheduling was essential. Using an online scheduling tool (e.g., Calendly) allowed participants to easily reschedule sessions as needed, reducing attrition and accommodating diverse work hours. 
Logistically, maintaining consistent procedures and clear communication across sessions was critical for ensuring data comparability while allowing participants' understanding of the system to evolve. 
We also observed that participants' reflections became richer as familiarity grew, suggesting that later sessions are particularly valuable for capturing deeper insights into human-AI collaboration. 
Future researchers conducting longitudinal studies may benefit from protocols that explicitly account for engagement, scheduling flexibility, and reflection across time.
}

\subsection{Impact of Perceived UX Expertise (RQ2)}

Our study showed that having any CA is better than none, as the no-CA condition resulted in significantly fewer identified problems, fewer unique problems, and lower inter-rater reliability, as indicated by Fleiss' kappa. 
Moreover, not only is any CA better than none, but the \textit{expertise of the CA also matters}: we found that the experienced CA led to significantly higher trust and efficiency than the novice CA, particularly after participants gained experience with these tools. 
While this partially aligns with prior research showing that high AI accuracy leads to greater trust in domains like legal decision-making \cite{kahr_understanding_2024}, our study extends these findings by investigating a different domain (usability analysis) over a longer time horizon, and by examining how perceived AI expertise influences evaluators' analytic behaviors beyond trust.
We further explore the implications of our findings concerning the differences between novice and experienced CAs.

\subsubsection{Implications for Real-World Collaboration with the Experienced CA}

The overarching goal of usability testing is to identify usability problems and uncover opportunities for improvement \cite{moran_usability_2019}. 
However, the reliability of analysis results is hindered by the \textit{``evaluator effect,''} where different evaluators tend to uncover different usability problems \cite{hertzum_evaluator_1998, hertzum_evaluator_2001}. 
Prior work has demonstrated weak inter-rater reliability in heuristic evaluations \cite{white_weak_2011, smith_reliability_2021} and high variability between evaluators in usability tests \cite{rohrer_practical_2016}. 
While the evaluator effect suggests that some core UX methods may be unreliable, agreement is not the only goal of usability testing \cite{sauro_how_2018}. 
Completeness is also crucial, as identifying a diverse set of problems can lead to overall usability improvements.

Prior research recommends collaboration among multiple evaluators to leverage diverse perspectives and identify a broader range of usability problems \cite{sauro_how_2018, hertzum_evaluator_1998}. Yet surveys show that only 23\% \cite{folstad_analysis_2012} and 37\% \cite{kuang_merging_2022} of UX evaluators in industry actually collaborate on analyzing usability videos. 
In practice, this data means a single evaluator often analyzes multiple videos. 
Our findings suggest that \textbf{collaborating with an experienced CA can effectively provide diverse perspectives without compromising reliability}: the experienced CA achieved the highest problem coverage and inter-rater reliability, making it a valuable aid for evaluators.
To further enhance evaluator-CA collaboration, future tools could include configurable prompts that specify the AI's level of expertise. They could also adopt distributed-agent approaches, such as DesignGPT, where a human inputs requirements and collaborates with multiple AI ``employees'' (e.g., product managers, design directors) \cite{ding_designgpt_2023}. 
Such capabilities would allow UX professionals to efficiently leverage CAs with greater expertise, improving both the completeness and reliability of usability analyses.


\subsubsection{Implications for Real-World Collaboration with the Novice CA}


While the experienced CA led to better analytic performance and was generally preferred, the novice CA was still perceived as valuable, particularly for \textit{training purposes}. 
The UX field is rapidly evolving with the introduction of new tools, design philosophies, and most recently, the integration of AI and machine learning into UX practices \cite{ironhack_evolving_2023}. 
Research has shown that CAs can boost productivity, especially for novice users, by helping them bridge skill gaps and accelerate learning \cite{brynjolfsson_generative_2023}. 
Building on these insights, we propose leveraging the \textbf{novice CA as a training partner to support the development of emerging UX evaluators}, providing a low-stakes environment to practice, reflect, and refine their analytic skills.

This perspective aligns with prior research on the \textit{protégé effect}, which shows that adopting a mentoring role, such as guiding a novice, can enhance one's own learning and critical thinking \cite{chase_teachable_2009}. 
Similarly, previous studies have found that UX evaluators perceived AI-generated usability problems as incomplete and likened the AI to a ``junior colleague'' or ``new intern who needs to be monitored'' \cite{fan_human-ai_2022}. 
Our findings echo this dynamic: participants described the novice CA in similar terms and engaged more critically with its suggestions.
\rv{Recent work also highlights the potential benefits of prosocial attitudes toward AI agents, suggesting that helping AI can enhance human well-being \cite{zhu_benefits_2025}. 
Future research could build on this by refining the novice-AI framing and exploring how designs that invite humans to ''support'' AI might better meet human needs and foster well-being.}

A common fear surrounding AI is the belief that it will eventually match human capabilities and replace jobs, which has been dubbed the ``fear of obsolescence'' (FOBO) \cite{whiting_is_2023}. 
In contrast, our study demonstrated that over time, participants recognized both the capabilities and limitations of the CA, leading to a productive division of labor rather than displacement. 
\rv{Recent work comparing a Nielsen-heuristics-trained GPT with human UX experts also highlighted its difficulty handling ambiguous scenarios, underscoring the need for human oversight in complex cases \cite{thai_smarter_2025}.}
Likewise, we argue that human expertise remains essential and that novice CAs may serve as effective training partners, helping evaluators sharpen their skills.
For instance, P4, who had two years of UX experience, shared that participating in this multi-session study felt like a training program, improving her attention to detail and articulation of usability problems. 
Our data also showed higher ignore rates for novice CAs, suggesting that participants actively evaluated and filtered AI suggestions rather than accepting them by default. 
This process of identifying and correcting AI mistakes encouraged evaluators to reflect on their own reasoning, reinforcing critical thinking and domain expertise. 
\rv{Extending this idea, future work could explore \textit{deliberative AI} for usability analysis, in which humans and AI jointly deliberate to resolve conflicting perspectives in decision-making tasks \cite{ma_towards_2025}.}

Furthermore, the CA's expertise should be considered relative to that of the human evaluator, meaning that, to remain an effective training partner, the CA's expertise may need to be updated over time as the human evaluator's skills improve.
Adaptation, where the AI learns the human decision-making process and updates its behavior accordingly, has shown benefits in training contexts such as UX and product design \cite{tao_designweaver_2025, buendia-garcia_using_2025}, flight simulation \cite{delisle_intelligent_2022}, higher education \cite{sajja_artificial_2024}, and navigation tasks \cite{zhao_role_2022}. 
In usability analysis, AI could similarly adapt to evaluators' evolving skills by adjusting its recommendations based on real-time signals like user hesitation, time spent on tasks, or patterns in the feedback process. 
This adaptation could enhance the AI's ability to provide more tailored support as evaluators' expertise develops.

\subsection{Limitations and Future Work}

In this study, we simulated novice and experienced CAs to explore how perceived UX expertise influences evaluators' analysis behaviors. 
While our prompts were grounded in general UX knowledge and practices, they may not fully reflect the diversity of individual evaluator styles and capabilities. 
Although the novice CA showed potential as a training aid, further research is needed to determine whether such interactions lead to measurable improvements in usability analysis skills. 
\rv{While we aimed to reduce confounds related to tone or semantics by focusing on task-related expertise and enforcing a strict response format,} prior work shows that a CA's personality traits, such as sociability, enthusiasm, or critical tone, can influence user perceptions and behavior \cite{pal_what_2023, lessio_toward_2020, pradhan_hey_2021}. 
Future studies could investigate how personality design choices impact evaluator engagement and collaboration, further refining the role of CAs in usability analysis.

Participants also interacted with a single CA per video, which differs from real-world settings where multiple evaluators with diverse expertise could collaborate on the same project. 
Expanding on previous research of chatbots facilitating consensus-building in co-design \cite{shin_chatbots_2022}, future studies could investigate environments where multiple CAs, each with different areas of expertise, collaborate on the same video. 
This would provide insights into how mixed-expertise teams interact, negotiate findings, and influence each other's analysis strategies. 

Lastly, this study involved a limited number of participants. 
The gender distribution (10 females and 2 males) is roughly consistent with the gender breakdown in crowdsourced data on UX researchers (76.3\% female, 20.3\% male) \cite{payscale_ux_2024}, though potential gender bias remains a consideration. 
\rv{Individual differences in participants’ prior AI experience may also have shaped their behaviors during the study.}
Future research with larger, more diverse samples could help validate these findings and determine whether similar longitudinal patterns emerge.
\section{Conclusion}

This study explored how evaluator behaviors evolve when collaborating with AI in usability analysis. Through a multi-session, within-subjects study, we found that participants adapted their analysis strategies over time and leveraged novice and experienced CAs in distinct ways, leading to effective analyses and a clear division of work between humans and AI. 
The presence of CA assistance significantly outperformed no assistance, boosting both analysis efficiency and evaluator confidence. 
The experienced CA was preferred, increasing the reliability of results without sacrificing diversity. 
Ultimately, our study highlights the potential of AI in improving usability evaluations and fostering more efficient and effective human-AI partnerships over time.

\bibliographystyle{ACM-Reference-Format}
\bibliography{bibs/main}

\newpage
\appendix
\section{Appendix}

\begin{figure*}[h]
  \centering
  \resizebox{\textwidth}{!}{%
        \scalebox{1}{\includegraphics{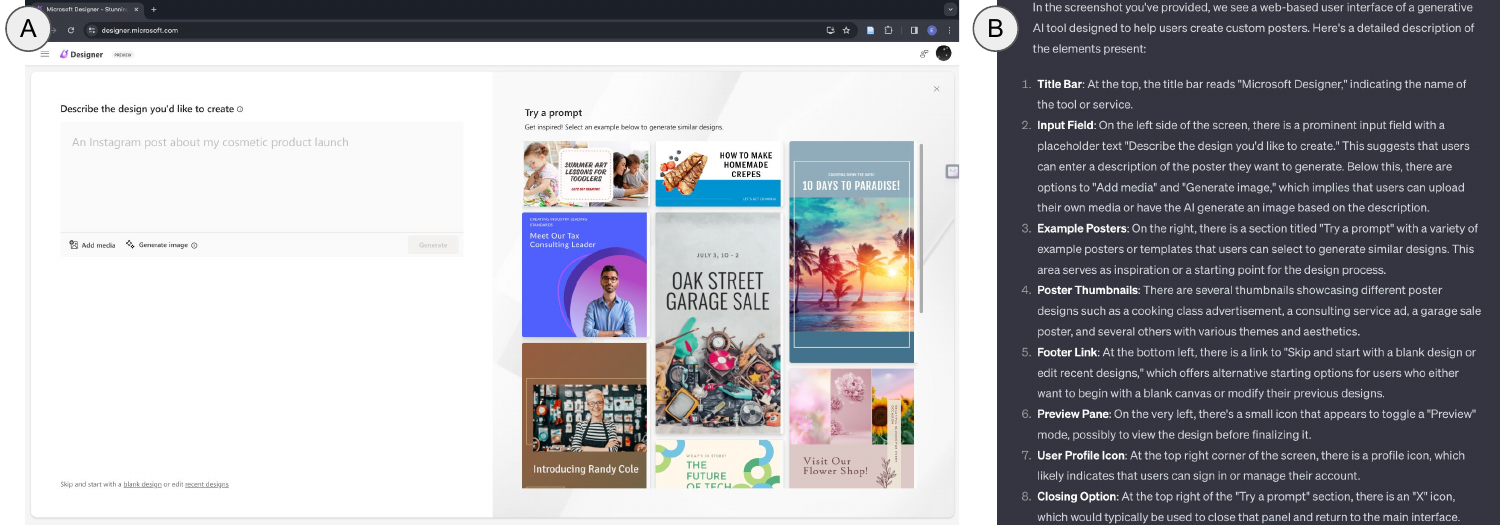}}
    }
  \caption{Screenshot of the homepage of a test website (A) and the corresponding response from GPT describing the UI elements shown in the screenshot (B).}
  \Description{Screenshot depicting two sections labeled 'A' and 'B.' Section A shows the test website interface with an input field asking to 'Describe the design you'd like to create' and several example posters to choose from on the right. Section B shows the response from GPT, which reads: In the screenshot you've provided, we see a web-based user interface of a generative AI tool designed to help users create custom posters. Here's a detailed description of the elements present: 1. Title Bar, 2. Input Field, 3. Example Posters, 4. Poster Thumbnails, 5. Footer Link, 6. Preview Pane, 7. User Profile Icon, 8. Closing Option}
  \label{fig-chp7:UI-understanding}
\end{figure*}

\begin{table}[h]
  \centering
  \caption{Prompts used to generate automatic suggestions and respond to questions}
  \label{tab-chp7:GPT-prompts}
  \small
  \begin{tabular}{p{1.5cm} p{6.2cm}}
    \toprule
    \textbf{Task} & \textbf{Prompt} \\
    \midrule
    Generate automatic suggestions & This is the transcript of a usability test where a participant used the think-aloud protocol to complete the following tasks: \textit{[insert video tasks from Table \ref{tab-chp7:videoInfo}]}
    \newline Transcript: \textit{[insert transcript from video]}
    \newline Based on your role as a \textit{[novice/experienced]} UX evaluator, as well as your knowledge of the user interfaces in the \textit{[insert product]}, please identify all the usability problems that the participant may have encountered, the timestamps when the problem occurred, the potential causes of why the problem occurred, and redesign recommendations to address the problem. Provide your response in the format: Problem description, Start time, End time, Cause of Problem, and Redesign recommendation \\
    \midrule
    Respond to questions &  This is the transcript of a usability test where a participant used the think-aloud protocol to complete the following tasks: \textit{[insert video tasks from Table \ref{tab-chp7:videoInfo}]}
    \newline Transcript: \textit{[insert transcript from video]}
    \newline This is the list of usability problems, causes, and redesign recommendations that you have previously identified in this video: \textit{[insert previously identified problem descriptions, causes, and redesign recommendations]}
    \newline Respond to the following questions based on your role as a \textit{[novice/experienced]}, as well as your knowledge of the user interfaces in the \textit{[insert product]}. Make your responses concise. \\
    \bottomrule
\end{tabular}
\end{table}

\begin{figure*}[h]
  \centering
  \resizebox{\textwidth}{!}{%
        \scalebox{1}{\includegraphics{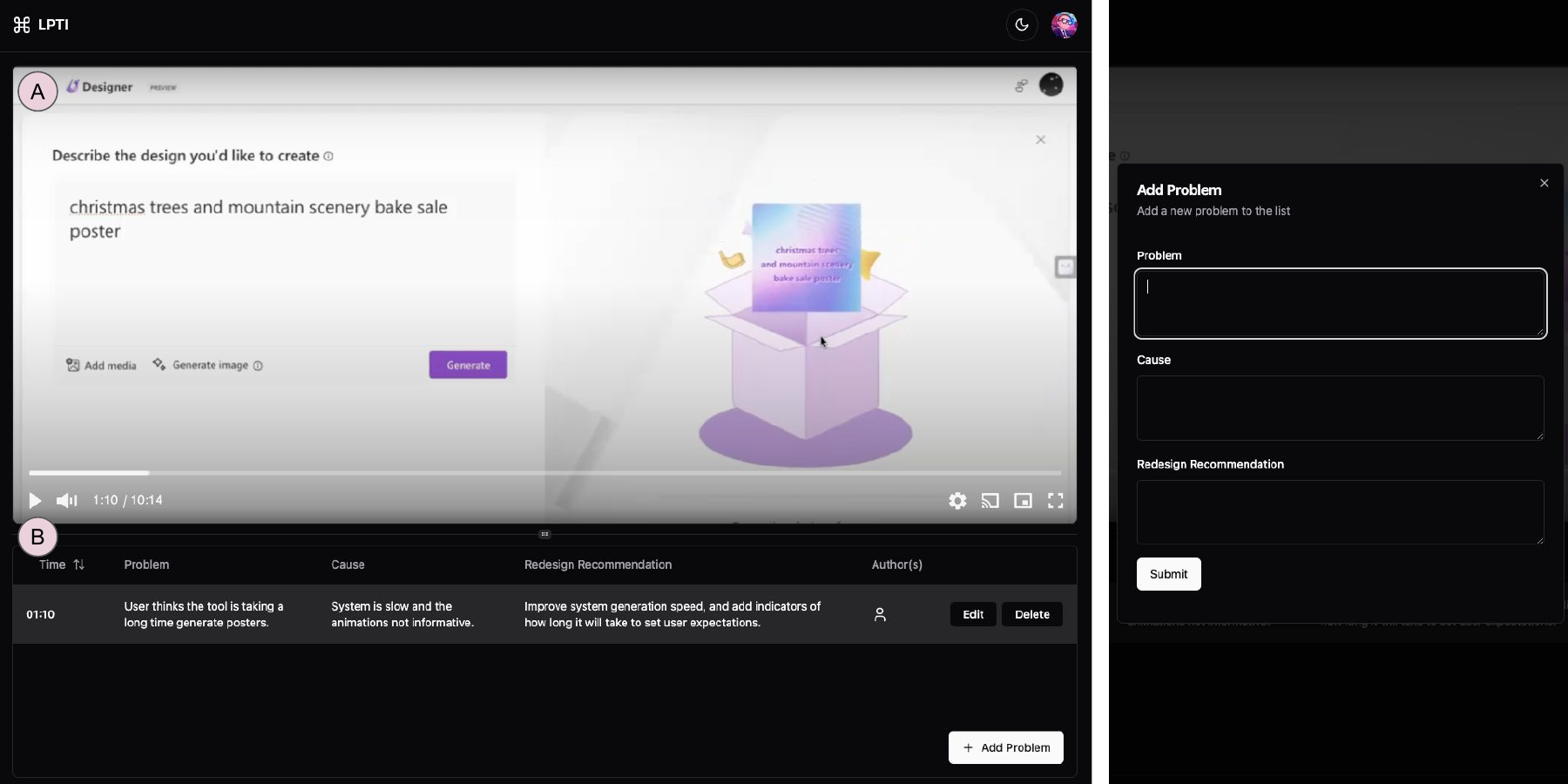}}
    }
  \caption{User interface of the UX analysis tool for the baseline condition containing: A) video player, and B) table of usability problem descriptions, causes, and redesign recommendations with the corresponding timestamp.}
  \Description{Screenshot of the user interface. Section A displays a video player paused at 1:10. Section B is a table listing a problem, cause, redesign recommendation, and author details for an issue noted at 1:10, which reads "Problem: User thinks the tool is taking a long time to generate posters. Cause: System is slow and the animations are not informative. Redesign Recommendation: Improve system generation speed, and add indicators of how long it will take to set user expectations. Author: User icon." There is an 'Add Problem' button at the bottom right corner of section B, and an empty modal form on the right to add a new problem to the list, with fields for Problem, Cause, and Redesign Recommendation.}
  \label{fig-chp7:prototype-UI-baseline}
\end{figure*}

\begin{figure*}[h]
  \centering
  \resizebox{\textwidth}{!}{%
        \scalebox{1}{\includegraphics{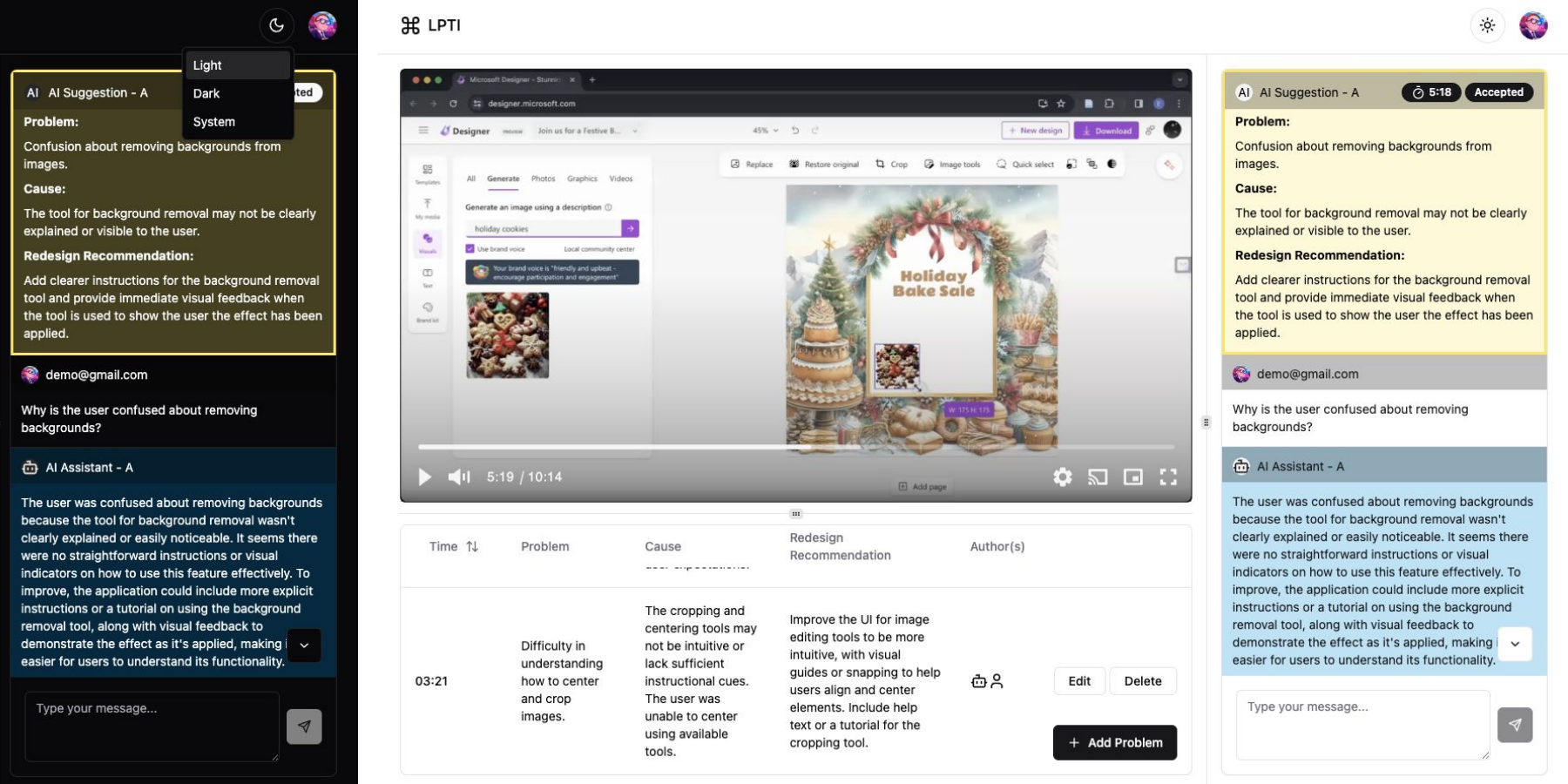}}
    }
  \caption{User interface of the UX analysis tool in light mode. Users can switch between light and dark modes by clicking on the ``moon'' or ``sun'' icon.}
  \Description{Screenshot of the UX analysis tool interface. The left section shows an AI suggestion panel in dark mode with a problem description about confusion over removing backgrounds from images. The right section, in light mode, mirrors the AI suggestion panel from the left with the same problem description.}
  \label{fig-chp7:prototype-UI-light}
\end{figure*}

\begin{table*}[h]
  \centering
  \caption{Information on the 15 usability videos used in the study}
  \label{tab-chp7:videoInfo}
  \small
  \begin{tabular}{p{1cm} p{7.5cm} p{1.5cm} p{1.5cm} p{2cm}}
    \toprule
    \textbf{Product} & \textbf{Tasks} & \textbf{Length (m:ss)} & \textbf{Number of \newline Suggestions (Novice)}  & \textbf{Number of \newline Suggestions \newline (Experienced)} \\
    \midrule
    Weather \newline App & 
    1. Customize the home screen so that UV is visible at the top. \newline
    2. Add Copenhagen to the list of saved locations and change the nickname of this city to ``Vacation.'' \newline
    3. Find the weather forecast for Copenhagen on Oct 16. \newline
    4. Remove the nickname ``Vacation'' and reset it to the default name.
    & 7:27 \newline 14:15 \newline 7:56 \newline 7:29 \newline 11:08 
    & 4 \newline 6 \newline 5 \newline 7 \newline 5
    & 6 \newline 6 \newline 6 \newline 9 \newline 7 \\
    \midrule
    GenAI \newline Website & 
    1. Create a poster to advertise a fundraising bake sale in a local community center that includes the date, time, and location of the bake sale and at least 3 different visualizations. \newline
    2. Export the poster, then generate a catchy caption and hashtags to post on Instagram.
    & 18:08 \newline 10:14 \newline 13:49 \newline 17:07 \newline 18:00
    & 6 \newline 7 \newline 6 \newline 7 \newline 9 
    & 8 \newline 9 \newline 8 \newline 9 \newline 13 \\
    \midrule
    VR Game & 
    1. Select the ``Blocking'' game from the menu and play it. \newline
    2. Select the ``Volleyball'' game from the menu and play it. \newline
    3. Select the ``Skeet'' game from the menu and play it. 
    & 7:51 \newline 9:42 \newline 10:56 \newline 9:06 \newline 11:27
    & 5 \newline 5 \newline 7 \newline 6 \newline 6
    & 7 \newline 8 \newline 7 \newline 7 \newline 6 \\
    \bottomrule
\end{tabular}
\end{table*}

\begin{table*}[h]
  \centering
  \caption{Categories of Messages sent by participants.}
  \label{tab:message-cateogories}
  \small
  \begin{tabular}{p{1.5cm} p{3.5cm} p{1.5cm} p{7cm}}
    \toprule
    \textbf{Session \newline Appearance} & \textbf{Category} & \textbf{Frequency (\%)} & \textbf{Example Question} \\
    \midrule
    \multirow{1}{1.5cm}{All sessions} & Problem identification & 39	(22.9\%) & ''Do you think there is a usability issue here at this timestamp?'' \\
    \midrule
    \multirow{8}{1.5cm}{Sessions 1 to 3} & 
    Instructions for AI & 11 (6.5\%) & ``Just wanted to let you know you should report each and every minor and major usability issue to me.'' \\
    & Task information & 10 (5.9\%) & ``What tasks did they complete?'' \\
    & AI capabilities & 7 (4.1\%) & ``Can you generate content for a usability problem, cause, and suggestion if I give you some basic information?'' \\
    & Product information & 6 (3.5\%) & ``Where is the UV placed on the app?'' \\
    & Timestamp information & 6 (3.5\%) & ``please list all the time stamps for usability problems.'' \\
    & User background & 2 (1.2\%) & ``information about the user'' \\
    \midrule
    \multirow{8}{1.5cm}{Sessions 3 to 5} & 
    Problem generation & 24	(14.1\%) & ``Write the problem, cause, redesign recommendation for users not knowing how to reload in the tutorial because they didn't see the demo first'' \\
    & Problem prioritization & 9 (5.3\%) & ``Prioritize the changes that we need to make to the app design to fix all of the usability issues found in this video'' \\
    & Suggestion correction & 8 (4.7\%) & ``It is also caused because important objects for the game are out of the user's point of view. A redesign option for this could also include having the user start facing the blocks.'' \\
    \midrule
    \multirow{10}{1.5cm}{Scattered throughout sessions 1 to 5} & 
    User actions & 12 (7.1\%) & ``How long did it take the user to find the right page to rename the location?'' \\
    & Social expressions & 8 (4.7\%) & ``cool thanks for helping'' \\
    & Suggestion clarification & 7 (4.1\%) & ``Why did you suggest this organization is the best way?''\\
    & User perceptions & 6 (3.5\%) & ``why is the user struggling with the game?'' \\
    & Heuristics identification & 5 (2.9\%) & ``Does this fall under the category of any heuristics?'' \\ 
    & Redesign recommendations & 5 (2.9\%) & ``Do you think renaming the buttons is also a good redesign recommendation?'' \\
    & Product response clarification & 5 (2.9\%) & ``How many times did the AI only give one image?'' \\
    \bottomrule
\end{tabular}
\end{table*}

\begin{table*}[h]
  \centering
  \caption{Results of ANOVA Tests for Quantitative Data (Significant effects are bolded)}
  \label{tab-chp7:stat-tests}
  \small
  \begin{tabular}{llll}
    \toprule
    \textbf{Variable} & \textbf{Session} & \textbf{Condition} & \textbf{Interaction Effect} \\
    \midrule
    Number of Pauses & 
    \textbf{$\boldsymbol{F_{4, 44} = 2.7, p < .05, \eta_{p}^{2} = 0.20}$} & 
    $F_{2, 22} = 1.6, p > .05, \eta_{p}^{2} = 0.13$ & 
    $F_{8, 88} = 0.3, p > .05, \eta_{p}^{2} = 0.02$ \\
    \midrule 
    Acceptance Rate &
    $F_{4, 44} = 1.3, p > .05, \eta_{p}^{2} = 0.11$ &
    \textbf{$\boldsymbol{F_{1, 11} = 9.9, p < .01, \eta_{p}^{2} = 0.47}$} &
    $F_{4, 44} = 1.2, p > .05, \eta_{p}^{2} = 0.1$ \\
    \midrule
    Edit Rate & 
    $F_{4, 44} = 0.7, p > .05, \eta_{p}^{2} = 0.06$ &
    $F_{1, 11} = 1.0, p > .05, \eta_{p}^{2} = 0.08$ & 
    $F_{4, 44} = 2.5, p > .05, \eta_{p}^{2} = 0.18$ \\
    \midrule
    Ignore Rate &
    \textbf{$\boldsymbol{F_{4, 44} = 3.1, p < .05, \eta_{p}^{2} = 0.22}$} &
    \textbf{$\boldsymbol{F_{1, 11} = 24.4, p < .001, \eta_{p}^{2} = 0.69}$} &
    $F_{4, 44} = 0.8, p > .05, \eta_{p}^{2} = 0.07$ \\
    \midrule
    Number of Total Problems & 
    \textbf{$\boldsymbol{F_{4, 44} = 3.1, p < .05, \eta_{p}^{2} = 0.22}$} & 
    \textbf{$\boldsymbol{F_{2, 22} = 13.3, p < .001, \eta_{p}^{2} = 0.55}$} & 
    $F_{8, 88} = 0.4, p > .05, \eta_{p}^{2} = 0.04$ \\
    \midrule
    Number of Unique Problems & 
    $F_{4, 8} = 0.5, p > .05, \eta_{p}^{2} = 0.21$ & 
    \textbf{$\boldsymbol{F_{2, 4} = 9.1, p < .05, \eta_{p}^{2} = 0.82}$} & 
    $F_{8, 16} = 0.8, p > .05, \eta_{p}^{2} = 0.30$ \\
    \midrule
    Problem Coverage & 
    $F_{4, 8} = 0.6, p > .05, \eta_{p}^{2} = 0.24$ & 
    \textbf{$\boldsymbol{F_{2, 4} = 9.1, p < .05, \eta_{p}^{2} = 0.82}$} & 
    $F_{8, 16} = 0.8, p > .05, \eta_{p}^{2} = 0.28$ \\
    \midrule 
    Fleiss' Kappa & 
    $F_{4, 8} = 0.7, p > .05, \eta_{p}^{2} = 0.25$ & 
    \textbf{$\boldsymbol{F_{2, 4} = 9.1, p < .05, \eta_{p}^{2} = 0.82}$} & 
    $F_{8, 16} = 0.4, p > .05, \eta_{p}^{2} = 0.18$ \\
    \midrule
    Perceived Efficiency &
    \textbf{$\boldsymbol{F_{4, 44} = 5.9, p < .01, \eta_{p}^{2} = 0.35}$} & 
    \textbf{$\boldsymbol{F_{1, 11} = 12.8, p < .01, \eta_{p}^{2} = 0.54}$} &
    $F_{4, 44} = 2.2, p > .05, \eta_{p}^{2} = 0.17$ \\
    \midrule
    Trust &
    \textbf{$\boldsymbol{F_{4, 44} = 4.2, p < .01, \eta_{p}^{2} = 0.27}$} & 
    \textbf{$\boldsymbol{F_{1, 11} = 16.7, p < .01, \eta_{p}^{2} = 0.60}$} &
    $F_{4, 44} = 0.4, p > .05, \eta_{p}^{2} = 0.04$ \\
    \midrule
    Perceived Completeness &
    $F_{4, 44} = 1.2, p > .05, \eta_{p}^{2} = 0.10$ & 
    \textbf{$\boldsymbol{F_{1, 11} = 14.2, p < .01, \eta_{p}^{2} = 0.56}$} &
    $F_{4, 44} = 1.6, p > .05, \eta_{p}^{2} = 0.12$ \\
    \midrule
    Confidence &
    $F_{4, 44} = 0.48, p > .05, \eta_{p}^{2} = 0.04$ & 
    \textbf{$\boldsymbol{F_{1, 11} = 6.0, p < .05, \eta_{p}^{2} = 0.35}$} &
    $F_{4, 44} = 0.49, p > .05, \eta_{p}^{2} = 0.04$ \\
    \midrule
    Perceived UX Experience &
    \textbf{$\boldsymbol{F_{4, 44} = 2.7, p < .05, \eta_{p}^{2} = 0.20}$} & 
    \textbf{$\boldsymbol{F_{1, 11} = 24.4, p < .001, \eta_{p}^{2} = 0.69}$} &
    \textbf{$\boldsymbol{F_{4, 44} = 3.1, p < .05, \eta_{p}^{2} = 0.22}$} \\
    \midrule
    Overall Satisfaction & 
    N/A & 
    \textbf{$\boldsymbol{F_{1, 11} = 12.3, p < .01, \eta_{p}^{2} = 0.53}$} & 
    N/A \\
    \bottomrule
\end{tabular}
\end{table*}

\begin{figure*}[h]
  \centering
  \resizebox{0.6\textwidth}{!}{%
        \scalebox{1}{\includegraphics{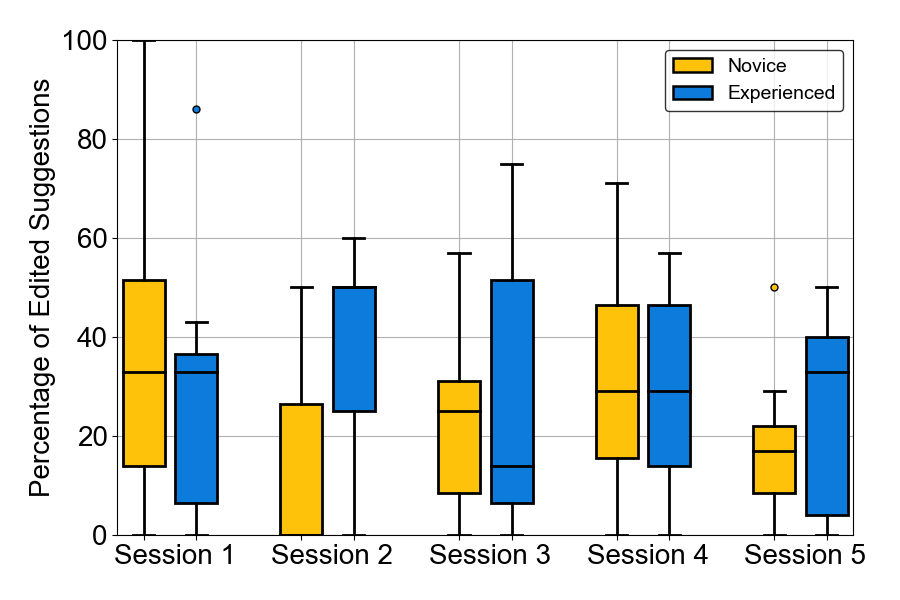}}
    }
  \caption{Box plot showing the percentage of edited suggestions separated by session and CA condition.}
  \Description{Box plot containing a y-axis for the percentage of suggestions, ranging from 0 to 100, and an x-axis with five pairs labeled from Session 1 to Session 5. Each session has two conditions: Low and High. There are no significant trends across sessions or conditions.}
  \label{fig-chp7:edit-rates}
\end{figure*}

\end{document}